\documentclass[longauth]{aa}

\usepackage{graphicx}
\usepackage{subcaption}
\usepackage{natbib}
\usepackage{scalerel}
\usepackage{booktabs}

\usepackage[table]{xcolor}

\usepackage{txfonts}
\usepackage[pdfencoding=auto,psdextra]{hyperref}

\usepackage{adjustbox}

\hypersetup{
    colorlinks=true,
    linkcolor=blue,
    filecolor=magenta,
    urlcolor=blue,
    citecolor=blue
}
\makeatletter
\renewcommand*\aa@pageof{, page \thepage{} of \pageref*{LastPage}}

\makeatother

\usepackage[utf8]{inputenc}

\usepackage[switch, modulo]{lineno}

\renewcommand{\linenumbers}[0]{}
\usepackage{euclid}
\usepackage[normalem]{ulem}

\newif\ifSGmark
\SGmarkfalse
\definecolor{SGnew}{RGB}{0,90,200}
\ifSGmark
  \newcommand{\chg}[1]{{\color{SGnew}#1}}
  \newcommand{\del}[1]{{\color{red}\sout{#1}}}
  \newcommand{\SGnote}[1]{{\color{magenta}\textbf{[SG: #1]}}}
\else
  \newcommand{\chg}[1]{#1}
  \newcommand{\del}[1]{}
  \newcommand{\SGnote}[1]{}
\fi
\providecommand{\logten}{\log_{10}}
\providecommand{\micron}{\ensuremath{\mu{\rm m}}}
\providecommand{\GN}{\ensuremath{G_{\rm N}}}

\providecommand{\AckDRone}{\SGnote{\textbackslash AckDRone undefined: update euclid.sty}}

\begin{document}
%
%

\title{\Euclid: Data Release 1 (DR1) -- A deep near-infrared search for low-mass and substellar companions to white dwarfs\thanks{This paper is published on behalf of the Euclid Consortium.}}    




\newcommand{\orcid}[1]{} 
\author{X.~P{\'e}rez-Couto\orcid{0000-0001-5797-252X}\thanks{\email{xabier.perez.couto@udc.gal}}\inst{\ref{aff1},\ref{aff2}}
\and A.~\v{Z}upi\'c\orcid{0009-0002-0811-2931}\inst{\ref{aff2},\ref{aff1}}
\and M.~Fern\'andez-Torreiro\orcid{0000-0002-6805-9100}\inst{\ref{aff2}}
\and S.~L.~Casewell\orcid{0000-0003-2478-0120}\inst{\ref{aff3}}
\and M.~Manteiga\orcid{0000-0002-7711-5581}\inst{\ref{aff2}}
\and C.~Dafonte\orcid{0000-0003-4693-7555}\inst{\ref{aff2},\ref{aff1}}
\and N.~Blagorodnova\orcid{0000-0003-0901-1606}\inst{\ref{aff4},\ref{aff5},\ref{aff6}}
\and B.~Altieri\orcid{0000-0003-3936-0284}\inst{\ref{aff7}}
\and S.~Andreon\orcid{0000-0002-2041-8784}\inst{\ref{aff8}}
\and N.~Auricchio\orcid{0000-0003-4444-8651}\inst{\ref{aff9}}
\and C.~Baccigalupi\orcid{0000-0002-8211-1630}\inst{\ref{aff10},\ref{aff11},\ref{aff12},\ref{aff13}}
\and M.~Baldi\orcid{0000-0003-4145-1943}\inst{\ref{aff14},\ref{aff9},\ref{aff15}}
\and A.~Balestra\orcid{0000-0002-6967-261X}\inst{\ref{aff16}}
\and S.~Bardelli\orcid{0000-0002-8900-0298}\inst{\ref{aff9}}
\and P.~Battaglia\orcid{0000-0002-7337-5909}\inst{\ref{aff9}}
\and A.~Biviano\orcid{0000-0002-0857-0732}\inst{\ref{aff11},\ref{aff10}}
\and E.~Branchini\orcid{0000-0002-0808-6908}\inst{\ref{aff17},\ref{aff18},\ref{aff8}}
\and M.~Brescia\orcid{0000-0001-9506-5680}\inst{\ref{aff19},\ref{aff20}}
\and V.~Capobianco\orcid{0000-0002-3309-7692}\inst{\ref{aff21}}
\and C.~Carbone\orcid{0000-0003-0125-3563}\inst{\ref{aff22}}
\and M.~Castellano\orcid{0000-0001-9875-8263}\inst{\ref{aff23}}
\and G.~Castignani\orcid{0000-0001-6831-0687}\inst{\ref{aff9}}
\and S.~Cavuoti\orcid{0000-0002-3787-4196}\inst{\ref{aff20},\ref{aff24}}
\and A.~Cimatti\inst{\ref{aff25}}
\and C.~Colodro-Conde\inst{\ref{aff26}}
\and G.~Congedo\orcid{0000-0003-2508-0046}\inst{\ref{aff27}}
\and L.~Conversi\orcid{0000-0002-6710-8476}\inst{\ref{aff28},\ref{aff7}}
\and Y.~Copin\orcid{0000-0002-5317-7518}\inst{\ref{aff29}}
\and F.~Courbin\orcid{0000-0003-0758-6510}\inst{\ref{aff4},\ref{aff30},\ref{aff31}}
\and H.~M.~Courtois\orcid{0000-0003-0509-1776}\inst{\ref{aff32}}
\and M.~Cropper\orcid{0000-0003-4571-9468}\inst{\ref{aff33}}
\and A.~Da~Silva\orcid{0000-0002-6385-1609}\inst{\ref{aff34},\ref{aff35}}
\and H.~Degaudenzi\orcid{0000-0002-5887-6799}\inst{\ref{aff36}}
\and G.~De~Lucia\orcid{0000-0002-6220-9104}\inst{\ref{aff11}}
\and H.~Dole\orcid{0000-0002-9767-3839}\inst{\ref{aff37},\ref{aff38}}
\and F.~Dubath\orcid{0000-0002-6533-2810}\inst{\ref{aff36}}
\and C.~A.~J.~Duncan\orcid{0009-0003-3573-0791}\inst{\ref{aff27}}
\and X.~Dupac\inst{\ref{aff7}}
\and S.~Dusini\orcid{0000-0002-1128-0664}\inst{\ref{aff39}}
\and M.~Fabricius\orcid{0000-0002-7025-6058}\inst{\ref{aff40},\ref{aff41}}
\and M.~Farina\orcid{0000-0002-3089-7846}\inst{\ref{aff42}}
\and R.~Farinelli\inst{\ref{aff9}}
\and F.~Faustini\orcid{0000-0001-6274-5145}\inst{\ref{aff23}}
\and S.~Ferriol\inst{\ref{aff29}}
\and F.~Finelli\orcid{0000-0002-6694-3269}\inst{\ref{aff9},\ref{aff43}}
\and N.~Fourmanoit\orcid{0009-0005-6816-6925}\inst{\ref{aff44}}
\and E.~Franceschi\orcid{0000-0002-0585-6591}\inst{\ref{aff9}}
\and M.~Fumana\orcid{0000-0001-6787-5950}\inst{\ref{aff22}}
\and L.~Gabarra\orcid{0000-0002-8486-8856}\inst{\ref{aff45}}
\and S.~Galeotta\orcid{0000-0002-3748-5115}\inst{\ref{aff11}}
\and K.~George\orcid{0000-0002-1734-8455}\inst{\ref{aff46}}
\and C.~Giocoli\orcid{0000-0002-9590-7961}\inst{\ref{aff9},\ref{aff15}}
\and J.~Gracia-Carpio\orcid{0000-0003-4689-3134}\inst{\ref{aff40}}
\and A.~Grazian\orcid{0000-0002-5688-0663}\inst{\ref{aff16}}
\and F.~Grupp\inst{\ref{aff40},\ref{aff41}}
\and S.~V.~H.~Haugan\orcid{0000-0001-9648-7260}\inst{\ref{aff47}}
\and S.~Hemmati\orcid{0000-0003-2226-5395}\inst{\ref{aff48}}
\and H.~Hoekstra\orcid{0000-0002-0641-3231}\inst{\ref{aff49}}
\and W.~Holmes\orcid{0009-0007-8554-4646}\inst{\ref{aff50}}
\and I.~M.~Hook\orcid{0000-0002-2960-978X}\inst{\ref{aff51}}
\and F.~Hormuth\inst{\ref{aff52}}
\and A.~Hornstrup\orcid{0000-0002-3363-0936}\inst{\ref{aff53},\ref{aff54}}
\and K.~Jahnke\orcid{0000-0003-3804-2137}\inst{\ref{aff55}}
\and M.~Jhabvala\inst{\ref{aff56}}
\and S.~Kermiche\orcid{0000-0002-0302-5735}\inst{\ref{aff44}}
\and B.~Kubik\orcid{0009-0006-5823-4880}\inst{\ref{aff29}}
\and M.~K\"ummel\orcid{0000-0003-2791-2117}\inst{\ref{aff41}}
\and M.~Kunz\orcid{0000-0002-3052-7394}\inst{\ref{aff57}}
\and H.~Kurki-Suonio\orcid{0000-0002-4618-3063}\inst{\ref{aff58},\ref{aff59}}
\and A.~M.~C.~Le~Brun\orcid{0000-0002-0936-4594}\inst{\ref{aff60}}
\and V.~Le~Brun\orcid{0000-0002-5027-1939}\inst{\ref{aff61}}
\and S.~Ligori\orcid{0000-0003-4172-4606}\inst{\ref{aff21}}
\and P.~B.~Lilje\orcid{0000-0003-4324-7794}\inst{\ref{aff47}}
\and V.~Lindholm\orcid{0000-0003-2317-5471}\inst{\ref{aff58},\ref{aff59}}
\and I.~Lloro\orcid{0000-0001-5966-1434}\inst{\ref{aff62}}
\and M.~Magliocchetti\orcid{0000-0001-9158-4838}\inst{\ref{aff42}}
\and G.~Mainetti\orcid{0000-0003-2384-2377}\inst{\ref{aff63}}
\and O.~Mansutti\orcid{0000-0001-5758-4658}\inst{\ref{aff11}}
\and O.~Marggraf\orcid{0000-0001-7242-3852}\inst{\ref{aff64}}
\and N.~Martinet\orcid{0000-0003-2786-7790}\inst{\ref{aff61}}
\and F.~Marulli\orcid{0000-0002-8850-0303}\inst{\ref{aff65},\ref{aff9},\ref{aff15}}
\and R.~J.~Massey\orcid{0000-0002-6085-3780}\inst{\ref{aff66}}
\and N.~Mauri\orcid{0000-0001-8196-1548}\inst{\ref{aff25},\ref{aff15}}
\and E.~Medinaceli\orcid{0000-0002-4040-7783}\inst{\ref{aff9}}
\and M.~Meneghetti\orcid{0000-0003-1225-7084}\inst{\ref{aff9},\ref{aff15}}
\and E.~Merlin\orcid{0000-0001-6870-8900}\inst{\ref{aff16}}
\and G.~Meylan\orcid{0000-0001-6503-0209}\inst{\ref{aff67}}
\and M.~Moresco\orcid{0000-0002-7616-7136}\inst{\ref{aff65},\ref{aff9}}
\and C.~Moretti\orcid{0000-0003-3314-8936}\inst{\ref{aff11},\ref{aff10},\ref{aff12}}
\and L.~Moscardini\orcid{0000-0002-3473-6716}\inst{\ref{aff65},\ref{aff9},\ref{aff15}}
\and R.~Nakajima\orcid{0009-0009-1213-7040}\inst{\ref{aff64}}
\and C.~Neissner\orcid{0000-0001-8524-4968}\inst{\ref{aff68},\ref{aff69}}
\and S.-M.~Niemi\orcid{0009-0005-0247-0086}\inst{\ref{aff70}}
\and C.~Padilla\orcid{0000-0001-7951-0166}\inst{\ref{aff68}}
\and S.~Paltani\orcid{0000-0002-8108-9179}\inst{\ref{aff36}}
\and F.~Pasian\orcid{0000-0002-4869-3227}\inst{\ref{aff11}}
\and W.~J.~Percival\orcid{0000-0002-0644-5727}\inst{\ref{aff71},\ref{aff72},\ref{aff73}}
\and V.~Pettorino\orcid{0000-0002-4203-9320}\inst{\ref{aff70}}
\and G.~Polenta\orcid{0000-0003-4067-9196}\inst{\ref{aff74}}
\and M.~Poncet\inst{\ref{aff75}}
\and L.~A.~Popa\inst{\ref{aff76}}
\and L.~Pozzetti\orcid{0000-0001-7085-0412}\inst{\ref{aff9}}
\and F.~Raison\orcid{0000-0002-7819-6918}\inst{\ref{aff40}}
\and A.~Renzi\orcid{0000-0001-9856-1970}\inst{\ref{aff77},\ref{aff39}}
\and J.~Rhodes\orcid{0000-0002-4485-8549}\inst{\ref{aff50}}
\and G.~Riccio\inst{\ref{aff20}}
\and F.~Rizzo\orcid{0000-0002-9407-585X}\inst{\ref{aff11}}
\and E.~Romelli\orcid{0000-0003-3069-9222}\inst{\ref{aff11}}
\and M.~Roncarelli\orcid{0000-0001-9587-7822}\inst{\ref{aff9}}
\and B.~Rusholme\orcid{0000-0001-7648-4142}\inst{\ref{aff48}}
\and R.~Saglia\orcid{0000-0003-0378-7032}\inst{\ref{aff41},\ref{aff40}}
\and Z.~Sakr\orcid{0000-0002-4823-3757}\inst{\ref{aff78},\ref{aff79},\ref{aff80}}
\and D.~Sapone\orcid{0000-0001-7089-4503}\inst{\ref{aff81}}
\and P.~Schneider\orcid{0000-0001-8561-2679}\inst{\ref{aff64}}
\and A.~Secroun\orcid{0000-0003-0505-3710}\inst{\ref{aff44}}
\and E.~Sihvola\orcid{0000-0003-1804-7715}\inst{\ref{aff58},\ref{aff59}}
\and P.~Simon\inst{\ref{aff64}}
\and C.~Sirignano\orcid{0000-0002-0995-7146}\inst{\ref{aff77},\ref{aff39}}
\and G.~Sirri\orcid{0000-0003-2626-2853}\inst{\ref{aff15}}
\and L.~Stanco\orcid{0000-0002-9706-5104}\inst{\ref{aff39}}
\and P.~Tallada-Cresp\'{i}\orcid{0000-0002-1336-8328}\inst{\ref{aff82},\ref{aff69}}
\and I.~Tereno\orcid{0000-0002-4537-6218}\inst{\ref{aff34},\ref{aff35}}
\and S.~Toft\orcid{0000-0003-3631-7176}\inst{\ref{aff83},\ref{aff84}}
\and R.~Toledo-Moreo\orcid{0000-0002-2997-4859}\inst{\ref{aff85},\ref{aff86}}
\and F.~Torradeflot\orcid{0000-0003-1160-1517}\inst{\ref{aff69},\ref{aff82}}
\and I.~Tutusaus\orcid{0000-0002-3199-0399}\inst{\ref{aff87},\ref{aff5},\ref{aff79}}
\and J.~Valiviita\orcid{0000-0001-6225-3693}\inst{\ref{aff58},\ref{aff59}}
\and T.~Vassallo\orcid{0000-0001-6512-6358}\inst{\ref{aff11},\ref{aff46}}
\and D.~Vibert\orcid{0009-0008-0607-631X}\inst{\ref{aff61}}
\and Y.~Wang\orcid{0000-0002-4749-2984}\inst{\ref{aff48}}
\and J.~Weller\orcid{0000-0002-8282-2010}\inst{\ref{aff41},\ref{aff40}}
\and G.~Zamorani\orcid{0000-0002-2318-301X}\inst{\ref{aff9}}
\and F.~M.~Zerbi\orcid{0000-0002-9996-973X}\inst{\ref{aff8}}
\and E.~Zucca\orcid{0000-0002-5845-8132}\inst{\ref{aff9}}
\and C.~M.~Gutierrez\orcid{0000-0001-7854-783X}\inst{\ref{aff26},\ref{aff88}}
\and K.~Tanidis\orcid{0000-0001-9843-5130}\inst{\ref{aff89}}
\and A.~Mohandasan\orcid{0000-0001-5182-0330}\inst{\ref{aff90}}}
										   
\institute{Universidade da Coru\~na (UDC), Department of Computer Science and Information Technologies, Campus de Elvi\~na s/n, 15071 A Coru\~na, Galiza, Spain\label{aff1}
\and
CIGUS CITIC, Centre for Information and Communications Technologies Research, Universidade da Coru\~na, Campus de Elvi\~na s/n, 15071 A Coru\~na, Spain\label{aff2}
\and
School of Physics and Astronomy, University of Leicester, University Road, Leicester, LE1 7RH, UK\label{aff3}
\and
Institut de Ci\`{e}ncies del Cosmos (ICCUB), Universitat de Barcelona (IEEC-UB), Mart\'{i} i Franqu\`{e}s 1, 08028 Barcelona, Spain\label{aff4}
\and
Institut d'Estudis Espacials de Catalunya (IEEC),  Edifici RDIT, Campus UPC, 08860 Castelldefels, Barcelona, Spain\label{aff5}
\and
Departament de F\'isica Qu\`antica i Astrof\'isica (FQA), Universitat de Barcelona (UB), Mart\'{\i} i Franqu\`es 1, E-08028 Barcelona, Spain\label{aff6}
\and
ESAC/ESA, Camino Bajo del Castillo, s/n., Urb. Villafranca del Castillo, 28692 Villanueva de la Ca\~nada, Madrid, Spain\label{aff7}
\and
INAF-Osservatorio Astronomico di Brera, Via Brera 28, 20122 Milano, Italy\label{aff8}
\and
INAF-Osservatorio di Astrofisica e Scienza dello Spazio di Bologna, Via Piero Gobetti 93/3, 40129 Bologna, Italy\label{aff9}
\and
IFPU, Institute for Fundamental Physics of the Universe, via Beirut 2, 34151 Trieste, Italy\label{aff10}
\and
INAF-Osservatorio Astronomico di Trieste, Via G. B. Tiepolo 11, 34143 Trieste, Italy\label{aff11}
\and
INFN, Sezione di Trieste, Via Valerio 2, 34127 Trieste TS, Italy\label{aff12}
\and
SISSA, International School for Advanced Studies, Via Bonomea 265, 34136 Trieste TS, Italy\label{aff13}
\and
Dipartimento di Fisica e Astronomia, Universit\`a di Bologna, Via Gobetti 93/2, 40129 Bologna, Italy\label{aff14}
\and
INFN-Sezione di Bologna, Viale Berti Pichat 6/2, 40127 Bologna, Italy\label{aff15}
\and
INAF-Osservatorio Astronomico di Padova, Via dell'Osservatorio 5, 35122 Padova, Italy\label{aff16}
\and
Dipartimento di Fisica, Universit\`a di Genova, Via Dodecaneso 33, 16146, Genova, Italy\label{aff17}
\and
INFN-Sezione di Genova, Via Dodecaneso 33, 16146, Genova, Italy\label{aff18}
\and
Department of Physics "E. Pancini", University Federico II, Via Cinthia 6, 80126, Napoli, Italy\label{aff19}
\and
INAF-Osservatorio Astronomico di Capodimonte, Via Moiariello 16, 80131 Napoli, Italy\label{aff20}
\and
INAF-Osservatorio Astrofisico di Torino, Via Osservatorio 20, 10025 Pino Torinese (TO), Italy\label{aff21}
\and
INAF-IASF Milano, Via Alfonso Corti 12, 20133 Milano, Italy\label{aff22}
\and
INAF-Osservatorio Astronomico di Roma, Via Frascati 33, 00078 Monteporzio Catone, Italy\label{aff23}
\and
INFN -- Sezione di Napoli, Via Cinthia 6, 80126, Napoli, Italy\label{aff24}
\and
Dipartimento di Fisica e Astronomia "Augusto Righi" - Alma Mater Studiorum Universit\`a di Bologna, Viale Berti Pichat 6/2, 40127 Bologna, Italy\label{aff25}
\and
Instituto de Astrof\'{\i}sica de Canarias, E-38205 La Laguna, Tenerife, Spain\label{aff26}
\and
Institute for Astronomy, University of Edinburgh, Royal Observatory, Blackford Hill, Edinburgh EH9 3HJ, UK\label{aff27}
\and
European Space Agency/ESRIN, Largo Galileo Galilei 1, 00044 Frascati, Roma, Italy\label{aff28}
\and
Universit\'e Lyon 1, CNRS/IN2P3, IP2I Lyon, Villeurbanne, France\label{aff29}
\and
Instituci\'o Catalana de Recerca i Estudis Avan\c{c}ats (ICREA), Passeig de Llu\'{\i}s Companys 23, 08010 Barcelona, Spain\label{aff30}
\and
Institut de Ciencies de l'Espai (IEEC-CSIC), Campus UAB, Carrer de Can Magrans, s/n Cerdanyola del Vall\'es, 08193 Barcelona, Spain\label{aff31}
\and
Universit\'e Lyon 1, CNRS/IN2P3, IUF, IP2I Lyon, Villeurbanne, France \label{aff32}
\and
Mullard Space Science Laboratory, University College London, Holmbury St Mary, Dorking, Surrey RH5 6NT, UK\label{aff33}
\and
Departamento de F\'isica, Faculdade de Ci\^encias, Universidade de Lisboa, Edif\'icio C8, Campo Grande, PT1749-016 Lisboa, Portugal\label{aff34}
\and
Instituto de Astrof\'isica e Ci\^encias do Espa\c{c}o, Faculdade de Ci\^encias, Universidade de Lisboa, Campo Grande, 1749-016 Lisboa, Portugal\label{aff35}
\and
Department of Astronomy, University of Geneva, ch. d'Ecogia 16, 1290 Versoix, Switzerland\label{aff36}
\and
Universit\'e Paris-Saclay, CNRS, Institut d'astrophysique spatiale, 91405, Orsay, France\label{aff37}
\and
Institut universitaire de France (IUF), 1 rue Descartes, 75231 PARIS CEDEX 05, France\label{aff38}
\and
INFN-Padova, Via Marzolo 8, 35131 Padova, Italy\label{aff39}
\and
Max Planck Institute for Extraterrestrial Physics, Giessenbachstr. 1, 85748 Garching, Germany\label{aff40}
\and
Universit\"ats-Sternwarte M\"unchen, Fakult\"at f\"ur Physik, Ludwig-Maximilians-Universit\"at M\"unchen, Scheinerstr.~1, 81679 M\"unchen, Germany\label{aff41}
\and
INAF-Istituto di Astrofisica e Planetologia Spaziali, via del Fosso del Cavaliere, 100, 00100 Roma, Italy\label{aff42}
\and
INFN-Bologna, Via Irnerio 46, 40126 Bologna, Italy\label{aff43}
\and
Aix-Marseille Universit\'e, CNRS/IN2P3, CPPM, Marseille, France\label{aff44}
\and
Department of Physics, University of Oxford, Keble Road, Oxford OX1 3RH, UK\label{aff45}
\and
University Observatory, LMU Faculty of Physics, Scheinerstr.~1, 81679 Munich, Germany\label{aff46}
\and
Institute of Theoretical Astrophysics, University of Oslo, P.O. Box 1029 Blindern, 0315 Oslo, Norway\label{aff47}
\and
Caltech/IPAC, 1200 E. California Blvd., Pasadena, CA 91125, USA\label{aff48}
\and
Leiden Observatory, Leiden University, Einsteinweg 55, 2333 CC Leiden, The Netherlands\label{aff49}
\and
Jet Propulsion Laboratory, California Institute of Technology, 4800 Oak Grove Drive, Pasadena, CA, 91109, USA\label{aff50}
\and
School of Physics and Astronomy, Lancaster University, Lancaster, LA1 4YB, UK\label{aff51}
\and
Felix Hormuth Engineering, Goethestr. 17, 69181 Leimen, Germany\label{aff52}
\and
Technical University of Denmark, Elektrovej 327, 2800 Kgs. Lyngby, Denmark\label{aff53}
\and
Cosmic Dawn Center (DAWN), Denmark\label{aff54}
\and
Max-Planck-Institut f\"ur Astronomie, K\"onigstuhl 17, 69117 Heidelberg, Germany\label{aff55}
\and
NASA Goddard Space Flight Center, Greenbelt, MD 20771, USA\label{aff56}
\and
Universit\'e de Gen\`eve, D\'epartement de Physique Th\'eorique and Centre for Astroparticle Physics, 24 quai Ernest-Ansermet, CH-1211 Gen\`eve 4, Switzerland\label{aff57}
\and
Department of Physics, P.O. Box 64, University of Helsinki, 00014 Helsinki, Finland\label{aff58}
\and
Helsinki Institute of Physics, Gustaf H{\"a}llstr{\"o}min katu 2, University of Helsinki, 00014 Helsinki, Finland\label{aff59}
\and
Laboratoire d'etude de l'Univers et des phenomenes eXtremes, Observatoire de Paris, Universit\'e PSL, Sorbonne Universit\'e, CNRS, 92190 Meudon, France\label{aff60}
\and
Aix-Marseille Universit\'e, CNRS, CNES, LAM, Marseille, France\label{aff61}
\and
SKAO, Jodrell Bank, Lower Withington, Macclesfield SK11 9FT, UK\label{aff62}
\and
Centre de Calcul de l'IN2P3/CNRS, 21 avenue Pierre de Coubertin 69627 Villeurbanne Cedex, France\label{aff63}
\and
Universit\"at Bonn, Argelander-Institut f\"ur Astronomie, Auf dem H\"ugel 71, 53121 Bonn, Germany\label{aff64}
\and
Dipartimento di Fisica e Astronomia "Augusto Righi" - Alma Mater Studiorum Universit\`a di Bologna, via Piero Gobetti 93/2, 40129 Bologna, Italy\label{aff65}
\and
Department of Physics, Institute for Computational Cosmology, Durham University, South Road, Durham, DH1 3LE, UK\label{aff66}
\and
Institute of Physics, Laboratory of Astrophysics, Ecole Polytechnique F\'ed\'erale de Lausanne (EPFL), Observatoire de Sauverny, 1290 Versoix, Switzerland\label{aff67}
\and
Institut de F\'{i}sica d'Altes Energies (IFAE), The Barcelona Institute of Science and Technology, Campus UAB, 08193 Bellaterra (Barcelona), Spain\label{aff68}
\and
Port d'Informaci\'{o} Cient\'{i}fica, Campus UAB, C. Albareda s/n, 08193 Bellaterra (Barcelona), Spain\label{aff69}
\and
European Space Agency/ESTEC, Keplerlaan 1, 2201 AZ Noordwijk, The Netherlands\label{aff70}
\and
Waterloo Centre for Astrophysics, University of Waterloo, Waterloo, Ontario N2L 3G1, Canada\label{aff71}
\and
Department of Physics and Astronomy, University of Waterloo, Waterloo, Ontario N2L 3G1, Canada\label{aff72}
\and
Perimeter Institute for Theoretical Physics, Waterloo, Ontario N2L 2Y5, Canada\label{aff73}
\and
Space Science Data Center, Italian Space Agency, via del Politecnico snc, 00133 Roma, Italy\label{aff74}
\and
Centre National d'Etudes Spatiales -- Centre spatial de Toulouse, 18 avenue Edouard Belin, 31401 Toulouse Cedex 9, France\label{aff75}
\and
Institute of Space Science, Str. Atomistilor, nr. 409 M\u{a}gurele, Ilfov, 077125, Romania\label{aff76}
\and
Dipartimento di Fisica e Astronomia "G. Galilei", Universit\`a di Padova, Via Marzolo 8, 35131 Padova, Italy\label{aff77}
\and
Instituto de F\'isica Te\'orica UAM-CSIC, Campus de Cantoblanco, 28049 Madrid, Spain\label{aff78}
\and
Institut de Recherche en Astrophysique et Plan\'etologie (IRAP), Universit\'e de Toulouse, CNRS, UPS, CNES, 14 Av. Edouard Belin, 31400 Toulouse, France\label{aff79}
\and
Universit\'e St Joseph; Faculty of Sciences, Beirut, Lebanon\label{aff80}
\and
Departamento de F\'isica, FCFM, Universidad de Chile, Blanco Encalada 2008, Santiago, Chile\label{aff81}
\and
Centro de Investigaciones Energ\'eticas, Medioambientales y Tecnol\'ogicas (CIEMAT), Avenida Complutense 40, 28040 Madrid, Spain\label{aff82}
\and
Cosmic Dawn Center (DAWN)\label{aff83}
\and
Niels Bohr Institute, University of Copenhagen, Jagtvej 128, 2200 Copenhagen, Denmark\label{aff84}
\and
Universidad Polit\'ecnica de Cartagena, Departamento de Electr\'onica y Tecnolog\'ia de Computadoras,  Plaza del Hospital 1, 30202 Cartagena, Spain\label{aff85}
\and
European University of Technology EUt+, European Union\label{aff86}
\and
Institute of Space Sciences (ICE, CSIC), Campus UAB, Carrer de Can Magrans, s/n, 08193 Barcelona, Spain\label{aff87}
\and
Universidad de La Laguna, Dpto. Astrof\'\i sica, E-38206 La Laguna, Tenerife, Spain\label{aff88}
\and
Center for Astrophysics and Cosmology, University of Nova Gorica, Nova Gorica, Slovenia\label{aff89}
\and
Universite Marie et Louis Pasteur, CNRS, Observatoire des Sciences de l'Univers THETA Franche-Comte Bourgogne, Institut UTINAM, Observatoire de Besan\c con, BP 1615, 25010 Besan\c con Cedex, France\label{aff90}}    

%
%
%
%


%
%
\abstract
{Low-mass and substellar companions to white dwarfs (WDs) are key to understanding binary evolution and irradiated substellar atmospheres. However, their identification through near-infrared (NIR) excesses has been limited by the depth and angular resolution of previous surveys.
We exploit Euclid DR1, which reaches about three magnitudes deeper than previous wide-area NIR surveys, at sub-arcsecond resolution, to carry out a homogeneous search for unresolved cool companions to WDs.
We cross-matched the Euclid DR1 MER catalogue with \textit{Gaia} to build a sample of 13\,122 high-confidence WDs in the Euclid Wide Survey. We anchored model atmospheres to the \textit{Gaia} photometry and identified significant flux excesses in the \JE and \HE bands using a magnitude-dependent significance threshold. We screened the candidates for contamination by resolved neighbours, fitted their WD-subtracted NISP slitless spectra with ultracool dwarf templates, and searched archival light curves from 12 surveys for periodic variability.
We identify 300 WDs with $3\sigma$ NIR excesses, a fraction of $2.6\%$; 107 of them are clean against resolved neighbours. The NISP spectra allowed us to characterise cool companions in 18 systems, with spectral types from M1 to T1, showing clear H$_2$O absorption in the later types. These include four brown dwarfs: three mid-to-late L dwarfs and a T dwarf. Three sources with strong Pa\,$\beta$ emission lines were also identified. One of these is a known period bouncer, and the other two are new candidates. Six clean NIR-excess candidates show periodic variability, including an eclipsing binary with an orbital period of $3.60\,\mathrm{h}$ and a spectroscopically confirmed WD--M6 system with a $99\,\mathrm{min}$ photometric period. A seventh shows a tentative signal near $11.23\,\mathrm{min}$.
\Euclid opens the faint end of the WD companion population to systematic study. Our catalogue includes all NIR-excess candidates, with contamination and variability flags to guide spectroscopic and time-domain follow-up.}
%
%
%
%
   \titlerunning{White dwarfs with low-mass companions in Euclid DR1}
   \authorrunning{Pérez-Couto et al.}

   \maketitle
%
%
%
%

\section{\label{sec:Intro}Introduction}

The vast majority of stars end their lives as white dwarfs \citep[WDs;][]{Iben1997}. Since between $50\%$ and $70\%$ of the stars are in binaries \citep{
Raghavan2010, DeRosa2014, Sana2014}, a substantial fraction of WDs are expected to have companions with longer evolutionary timescales than those of their progenitors, such as M dwarfs \citep{Farihi2005}.

Because WD cooling ages can be predicted reliably, the identification of such WD plus M-dwarf systems (hereafter, WDMS) is critical to \chg{parameterise} the age\chg{--}metallicity \chg{relation} \citep{Mansergas2026}, \chg{the initial-to-final mass relation} \citep{Zhao2012},
and\chg{, if the system is eclipsing, the mass--radius relation} \citep{Raddi2025}. In wide binaries, WDs also provide independent ages for low-mass main-sequence companions, whose slow evolution makes conventional age determinations challenging. Their cooling ages, combined with progenitor lifetimes, offer valuable benchmarks for calibrating methods such as gyrochronology and asteroseismology \citep{Garces2011, RebassaMansergas2013}.

In contrast, systems of WD plus substellar companions such as \chg{L and T dwarfs} are extremely rare, consistent with the \chg{brown dwarf} (BD) desert observed around main-sequence (MS) stars \citep{grether2006}. Indeed, the substellar fraction around
\del{d} WDs is consistently found to be \chg{approximately 0.1--0.5\%} \citep{fbz2005, girven2011, steele2011}, and only a few dozen of such systems are known to date \citep[e.g.\chg{,}][]{Casewell2015, Casewell2018, Casewell2024}.

Despite their scarcity, WD–BD binaries are particularly valuable for understanding binary evolution. Of special interest are close systems that may have survived a common-envelope (CE) phase, leaving a \chg{WD} with a substellar companion in a compact orbit. Such post-common-envelope binaries (PCEBs) may eventually evolve into cataclysmic variables (CVs) through orbital angular momentum loss \citep{Hernandez2016, Rappaport2017}. Establishing a connection with this evolutionary channel requires information on the orbital periods and separations of the systems. Time-domain photometry provides a way to identify close systems through orbital variability. WD–BD binaries are also valuable laboratories for studying irradiated substellar atmospheres, offering useful analogues to hot Jupiters \citep{Casewell2015, french2023, Amaro2025, Broski2026}.

Because the \chg{spectral energy distribution} (SED) of WDs typically peaks in the \chg{ultraviolet (UV)} and falls steeply through the optical, a cool, self-luminous companion (M, L\chg{,} or \chg{T dwarf}) whose SED peaks at wavelengths longer than \chg{1\,\micron}\del{,} can dominate the combined flux beyond that limit, producing a characteristic excess in the \chg{NIR} \citep{probst1983, fbz2005, steele2011, debes2011, lai21}.

IR excess searches have therefore relied mainly on SED fitting: WD models are fitted to UV \citep{Bianchi2017} and optical \citep{Alam2015, Riello2021, Benitez2014} photometry, and the predicted IR flux is compared with that observed by surveys such as \chg{the Two Micron All Sky Survey \citep[2MASS;][]{Skrutskie2006}, the UKIRT Infrared Deep Sky Survey \citep[UKIDSS;][]{lawrence07}, and CatWISE2020 \citep{Marocco2021}; see, for example, \citet{murillo2026}, \citet{girven2011}, and \citet{steele2011}}.

This methodology applied to these large IR surveys enabled the identification of hundreds of WDMS candidates. However, while photometry readily flags \chg{an} NIR excess, disentangling a genuine companion from background contamination generally requires spectroscopic confirmation. Such data \chg{allow us} to study the absorption and emission features characteristic of each companion type.

Current limitations are largely instrumental. The all-sky 2MASS survey reaches only $J_{\rm AB}\approx16.7$ \citep{Skrutskie2006}, while the deeper ground-based UKIDSS and \chg{Visible and Infrared Survey Telescope for Astronomy (VISTA)} surveys reach $J_{\rm AB}\approx21$ at seeing-limited resolution \citep[\chg{about} 1\arcsec;][]{McMahon2013}. At longer wavelengths, the \chg{6\arcsec\ point-spread function (PSF) of the Wide-field Infrared Survey Explorer (WISE)} \citep{Wright2010} makes blending with background sources a dominant source of spurious excesses.

The \Euclid mission
\citep{Laureijs11, EuclidSkyOverview} transforms this landscape. Its \chg{1.2\,m} telescope
feeds two instruments: the Visible Imager \citep[VIS;][]{EuclidSkyVIS}, providing
high-resolution imaging in a single broad optical band $\IE$, and the
Near-Infrared Spectrometer and Photometer \citep[NISP;][]{EuclidSkyNISP,
Schirmer-EP18}, delivering low-resolution slitless spectra \citep[$R\approx 450$;][Euclid Collaboration: Scodeggio et al., in prep.]{EuclidSkyNISP, DR1-TP024, Q1-TP006} and photometry in three NIR bands, $\YE$ (937.5--1243.2\,nm), $\JE$ (1151.1--1595.0\,nm), and $\HE$ (1495.6--2056.8\,nm), to a $5\sigma$ depth of approximately 24 (AB). This is approximately 3 magnitudes deeper than the deepest wide-area ground-based NIR surveys and approximately 7 magnitudes deeper than 2MASS, at an angular resolution limited by the 0\farcs1 (VIS) and 0\farcs3 (NISP) pixels. No ground-based survey can match such resolution over a comparable area \citep[around 14\,000\,deg$^{2}$ at the end of the mission;][]{EuclidSkyOverview}. As an example, a mid-L dwarf companion at a distance of 300\,pc would have \JE$\approx 21.9$ and a mid-T dwarf \JE$\approx 23.4$ \citep{sanghi2024}, below the detection limits of all previous wide-area NIR surveys but well within reach of \textit{Euclid}.

The potential of these data for the cool, low-mass stellar and substellar regime has already been demonstrated with the Quick Data Release 1 (Q1), which yielded thousands of
new ultracool dwarf candidates in the Euclid Deep Fields \citep{Q1-SP042,
	Q1-SP061} and confirmed some of them with NISP spectra \citep{Dominguez2025}. The first Data Release \citep[DR1;][]{DR1cite, Q1-TP001}\SGnote{in-prep.\ entry euclidDR1 replaced by the DR1 DOI entry DR1cite from Euclid.bib (SG Sect.~3.4, 2.7.2)},
covering \chg{approximately} 1900\,$\mathrm{deg}^2$ of the Wide Survey, provides for the first
time deep, high-resolution NIR photometry for a large, untargeted sample of
\chg{\textit{Gaia}} WDs. Its high angular resolution allows us to identify IR excesses co-located with these WDs while minimising the chances of the emission coming from a background source. Furthermore, ground-based $J$ and $H$ filters, such as those of 2MASS and UKIDSS, are designed to avoid the telluric water-vapour bands, which coincide with the strongest H$_2$O absorption in BD atmospheres.

In this work we exploit Euclid DR1 to carry out a homogeneous search
for unresolved cool companions to WDs. \chg{Our aims are to (i) identify WDs
with a significant NIR excess over their photospheric emission and quantify
how frequent such excesses are; (ii) distinguish genuine unresolved companions
from contamination by resolved neighbouring sources; (iii) characterise the companions spectroscopically across the low-mass and substellar regimes; and (iv) identify close systems among them, which probe post-common-envelope evolution and irradiated substellar atmospheres.}

The paper is organised as follows. Section~\ref{sc:euclid_wd_sample} describes the \Euclid WD sample and its properties, \chg{Sect.}~\ref{sec:identificationirexcess} describes the model-atmosphere fitting, the photometric data, the identification of NIR excesses, and the assessment of contamination by resolved neighbours. Subsequently, in \chg{Sect.}~\ref{sec:nisp_spectra} we spectroscopically characterise the clean candidates using NISP spectra. In Sect.~\ref{sec:variability} we perform a variability analysis through external, time-domain photometry. Finally, \chg{Sect.~\ref{sec:conclusions}} summarises our conclusions.

\section{\Euclid WD sample \label{sc:euclid_wd_sample}}

\subsection{Sample selection}

We identified WDs in Euclid DR1 by cross-matching the \chg{merged (MER)} catalogue with the \chg{\textit{Gaia}} WD catalogue of \citet{gf2021}, which provides for each source a probability $P_{\rm WD}$ of being a WD based on its position in the \textit{Gaia} colour\chg{--}magnitude diagram\del{ (CMD)}. We required $P_{\rm WD}> 0.75$ to limit contamination by non-WDs, in addition to a positive parallax with a signal-to-noise \chg{ratio} (S/N) greater than 3, to ensure reliable distance estimates \citep{Weiler2025}.

Subsequently, we cross-matched this sample of $309\,003 $ \textit{Gaia} WDs with the \chg{Euclid DR1 MER} catalogue \citep{Q1-TP001}, requiring a detection in the \IE band (\texttt{vis\_det = 1}), with additional cuts to remove saturated sources (\texttt{det\_quality\_flag<3}) and negative fluxes. \chg{Appendix}~\ref{sec:adql_query} shows the \chg{Astronomical Data Query Language (ADQL) query} used in the \chg{Euclid} Science Archive\footnote{\url{https://eas.esac.esa.int/sas/}}.

The final sample comprises $N_\mathrm{WD}=13\,122$ \textit{Euclid} WDs. We corrected their \chg{\textit{Gaia}} parallaxes using \citet{Lindegren2021} and the proper motions following  \citet{Cantat-Gaudin2021}, and increased the parallax uncertainties following \citet{MaizApellaniz2022}. Subsequently, heliocentric distances were obtained following the methods in  \citet{Weiler2025}. The full catalogue of the 13\,122 \textit{Euclid} WDs is available at the CDS, and its columns are described in Table~\ref{tab:cds_columns}.

\subsection{Sample properties}

\begin{figure}[ht]
    \centering
    \includegraphics[width=0.5\textwidth]{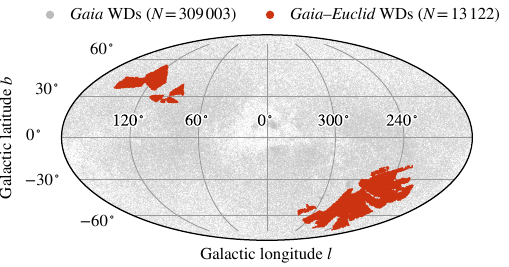}
    \caption{Sky distribution of the \Euclid WDs (in red), with the \chg{\textit{Gaia}} WDs (in grey) as background.}
    \label{fig:euclid_wd_sky}
\end{figure}

As noted in Sect. \ref{sec:Intro}, DR1 covers about 1900\,deg$^{2}$ of the Wide Survey ($\simeq5\%$ of the sky), which avoids the Galactic and ecliptic planes. Because of this partial sky coverage, it is important to compare our \Euclid WD sample with the all-sky \chg{\textit{Gaia}} parent sample in order to identify possible observational bias. In \chg{Fig.}~\ref{fig:euclid_wd_sky} we show the sky distribution of the \Euclid WDs in comparison to the all-sky \chg{\textit{Gaia}} sample.

Furthermore, we compared the two samples across different parameters: apparent magnitude $G$, absolute magnitude $M_G$, \chg{$G_{\rm BP} - G_{\rm RP}$} colour, extinction $A_V$, distance $d$, effective temperature $T_{\rm eff}$, mass $M_{\rm WD}$, and surface gravity $\log{g}$ \chg{[we use $\log g \equiv \logten(g/{\rm cm\,s^{-2}})$ throughout]}, the last three coming from the \citet{gf2021} \chg{catalogue} assuming H-rich atmospheres (and available for \chg{$11\,924$} sources). For each parameter, Table \ref{tab:representativeness} lists the median and 16th--84th percentile range of each population. As effect sizes we give the median offset $\Delta_{\rm med}=\textit{Euclid}-\textit{Gaia}$, the Wasserstein distance $W_1$, and \chg{common language effect size} (CLES), with 16th--84th percentile uncertainties from 1000 bootstrap iterations. CLES is computed here as the Mann–Whitney $U$-statistic \chg{normalised} by the total number of pairs, and it provides the probability that a value drawn from the \textit{Gaia} sample exceeds one drawn from the \Euclid sample. The Wasserstein distance, on the other hand, provides the distance between two probability distributions, thus complementing $\Delta_{\rm med}$. If both samples were drawn from the same population, we would expect
$\Delta_{\rm med}$ to be consistent with zero, $W_1$ to be comparable to
its finite-sample noise, and CLES $\approx 0.5$. Given the large
sample sizes, however, even negligible differences can be statistically
significant, so we assess the size of the offsets rather than only their
significance.

The largest relative difference between both populations manifests in the extinction, $A_V$, as expected \chg{from} the \Euclid~\chg{sky footprint}, far away from the Galactic plane (\chg{Fig.}~\ref{fig:euclid_wd_sky}). Apart from that, the only appreciable difference is a modest shift towards fainter, more distant objects: the \Euclid sample is fainter by $\Delta_{\rm med}(G)=+0.10$ and more distant by $\Delta_{\rm med}(d)=+17.6$\,pc (a 6\% increase). The absolute-magnitude distributions are almost identical \chg{[}$\Delta_{\rm med}(M_G)=-0.02$, $W_1=0.05$\chg{]}, so the apparent faintness reflects distance. Indeed, the $0.10$ offset is consistent with the $6\%$ distance shift, once the lower extinction is included. The \Euclid sample is also marginally cooler \chg{[}$\Delta_{\rm med}(T_{\rm eff})=-510$\,K, \chg{approximately} 5\%\chg{]}, with correspondingly small offsets in mass and surface gravity \chg{[}$\Delta_{\rm med}(M_{\rm WD})=-0.02\,M_\odot$, $\Delta_{\rm med}(\log g)=-0.04$\chg{]}, consistent with a slight preference for cooler, lower-mass remnants at the few per cent level; the $G_{\rm BP} -G_{\rm RP}$ colour, $M_G$, $M_{\rm WD}$, and $\log g$ distributions are otherwise indistinguishable, and all CLES values lie within $0.46$ and $0.7$\SGnote{unclear: between 0.4 and 0.7?}. Hence, we conclude that the \Euclid sample traces the parent \chg{\textit{Gaia}} WD population without \chg{appreciable} bias in its intrinsic properties.

\begin{table*}
\caption{Properties of the \Euclid WDs relative to the parent \chg{\textit{Gaia}} sample.}
\label{tab:representativeness}
\centering
\renewcommand{\arraystretch}{1.5}
\begin{tabular}{lccccc}
\toprule
Parameter & Median (\chg{\textit{Gaia}}) & Median (\Euclid) & $\Delta_{\rm med}$ & $W_1$ & CLES \\ 
\midrule
$G$ & $19.72^{+0.62}_{-1.05}$ & $19.81^{+0.53}_{-0.83}$ & $+0.10^{+0.01}_{-0.01}$ & $0.20^{+0.01}_{-0.01}$ & $0.458$ \\
$M_G$ & $12.31^{+1.49}_{-1.24}$ & $12.29^{+1.52}_{-1.23}$ & $-0.02^{+0.02}_{-0.02}$ & $0.048^{+0.011}_{-0.001}$ & $0.502$ \\
\chg{$G_{\rm BP} - G_{\rm RP}$} & $0.158^{+0.444}_{-0.309}$ & $0.152^{+0.440}_{-0.294}$ & $-0.006^{+0.004}_{-0.004}$ & $0.011^{+0.003}_{-0.001}$ & $0.500$ \\
$A_V$ & $0.090^{+0.160}_{-0.060}$ & $0.050^{+0.050}_{-0.040}$ & $-0.040^{+0.005}_{-0.005}$ & $0.088^{+0.001}_{-0.001}$ & $0.700$ \\
$d$ [pc] & $280.4^{+194.6}_{-130.7}$ & $298.0^{+206.6}_{-134.2}$ & $+17.6^{+2.0}_{-2.6}$ & $20.8^{+1.7}_{-1.5}$ & $0.466$ \\
$T_{\rm eff}$ [K] & $10693^{+6804}_{-3923}$ & $10184^{+5476}_{-3527}$ & $-510^{+50}_{-51}$ & $1027^{+59}_{-57}$ & $0.536$ \\
$M_{\rm WD}$ [$M_{\odot}$] & $0.622^{+0.232}_{-0.177}$ & $0.598^{+0.222}_{-0.166}$ & $-0.024^{+0.002}_{-0.002}$ & $0.026^{+0.002}_{-0.002}$ & $0.536$ \\
$\log{g}$ & $8.029^{+0.367}_{-0.362}$ & $7.991^{+0.356}_{-0.360}$ & $-0.038^{+0.003}_{-0.004}$ & $0.046^{+0.004}_{-0.004}$ & $0.533$ \\
\bottomrule
\end{tabular}
\tablefoot{Medians are quoted with 16th--84th percentile ranges; the median offset $\Delta_{\rm med}=\chg{\mathrm{\Euclid}-\mathrm{\textit{Gaia}}}$ and the Wasserstein distance $W_1$ are effect sizes with bootstrap 16th--84th percentile uncertainties. $A_V$ is quantized in steps of 0.01, so the uncertainty on its
$\Delta_{\rm med}$ is set by this resolution ($\pm0.005$). For $M_G$ and
$G_{BP}-G_{RP}$, $W_1$ is comparable to the bootstrap noise floor.}
\end{table*}

\section{\texorpdfstring{Identification of \chg{NIR excesses}}{Identification of NIR excesses}}\label{sec:identificationirexcess}

In the following sections we use a subsample of \chg{11\,881} \Euclid WDs, drawn from the full sample defined in \chg{Sect.}~\ref{sc:euclid_wd_sample}, with available $T_{\rm eff}$ and $\log{g}$ in \citet{gf2021} and with $\mathrm{S/N} >5$ in all VIS and NIR bands.

 \subsection{White dwarf model fitting}\label{sec:wd_model_fitting}

To obtain the \chg{synthetic fluxes in the} $\IE$, $\YE$, $\JE$, \chg{and} $\HE$ \chg{bands} for each WD\chg{,} we used the grid of 1D pure-hydrogen \chg{non-local thermodynamic equilibrium (NLTE)} synthetic spectra of \cite{Tremblay2011}\footnote{\url{https://warwick.ac.uk/fac/sci/physics/research/astro/people/tremblay/modelgrids/}}, with \chg{$\log g\in[6.5, 9.5]$} and \chg{$T_{\rm eff}/{\rm K}\in[1500, 140\,000]$}. By interpolating in $T_{\rm eff}$ and $\log{g}$, we obtain model spectra in Eddington fluxes, \chg{$H_\nu$}.

Subsequently, we reddened the model spectra using the mean $A_V$ from \citet{gf2021} and the extinction curve model of \citet{Gordon2023}, based on \citet{Gordon2009}, \citet{Fitzpatrick2019}, \citet{Gordon2021}, and \citet{Decleir2022}, using the \texttt{dust\_extinction} package \citep{Gordon2024}.

The transmission curves of each band \citep{Rodrigo2020}\footnote{\url{https://svo2.cab.inta-csic.es/theory/fps/}} were used to integrate the flux and \chg{synthesise} the model fluxes\chg{, $F^{\rm mod}(b)$, in the \textit{Gaia} and \Euclid bands, $b \in \{G_{\rm BP}, G, G_{\rm RP}, \IE, \YE, \JE, \HE\}$}.

Finally, we transformed \chg{the} Eddington fluxes \chg{$H_\nu$} to the fluxes at the Earth, \chg{$F^{\rm mod}$}. This is done by using \chg{$F^{\rm mod}=4\pi\,(R_{\rm WD}/d)^2 H_\nu$}, where \chg{$R_{\rm WD} = \sqrt{\GN M_{\rm WD}/g}$} is the WD radius. To avoid propagating systematics in the model-derived $M_{\rm WD}$ and $\log{g}$, we instead fitted $\alpha = (R_{\rm WD}/d)^2$ to the \textit{Gaia} photometry, which anchors the model for the NIR excess detection.

\subsection{\Euclid photometry}\label{sec:euclid_photometry}

The \Euclid MER pipeline provides fluxes of different \chg{kinds}, each of which may be useful depending on the type of source being studied. Aperture photometry is processed with the \texttt{A-PHOT} tool \citep{Merlin2019} and is available in different apertures in terms of the \chg{full width at half maximum} (FWHM). \chg{Template-fitting} photometry is obtained with \texttt{T-PHOT} \citep{Merlin2015, Merlin2016} using the morphology of the VIS image as prior, and it is available for the NIR bands ($\YE, \JE, \HE$). The combination of high-resolution VIS morphology with NISP photometry, obtained simultaneously and from space, is unique to \textit{Euclid}, and is what enables the neighbour analysis of Sect.~\ref{sec:photometric_contamination}.


Following the validation results of \citet{Q1-TP004} for stars, we chose the \texttt{A-PHOT} photometry. To avoid the coarse $4 \times \mathrm{FWHM}$ aperture, which would admit more light from neighbours and waste \textit{Euclid}’s resolution, we used the $2 \times \mathrm{FWHM}$ fluxes and corrected them to total fluxes with the $\IE$ aperture-to-total ratio \citep[equation~3 in][]{Q1-TP004}. The corresponding uncertainties are those provided by \texttt{A-PHOT}. The median \IE of the sample with 16th--84th percentiles is $20^{+0.5}_{-0.8}$.

\subsection{Selection of NIR excesses}\label{sec:nir_excesses}

We followed \citet{Wilson2019} to define \chg{the} flux excess in \chg{a band $b \in \{\IE, \YE, \JE, \HE\}$} in terms of its significance\del{:}
\begin{equation}\label{fluxexcess}
    \chg{\chi(b) = \frac{F^{\rm obs}(b) - F^{\rm mod}(b)}{\sqrt{\sigma^2\left[F^{\rm obs}(b)\right] + \sigma^2\left[F^{\rm mod}(b)\right]}}}\;,
\end{equation}
where \chg{$F^{\rm obs}(b)$ and $F^{\rm mod}(b)$} are the observed and model fluxes in \chg{band $b$}, and \chg{$\sigma[F^{\rm obs}(b)]$ and $\sigma[F^{\rm mod}(b)]$} their uncertainties, respectively.

We adopted a $5\%$ uncertainty on the model fluxes \citep{Wilson2019}. A large positive $\chi$ then indicates flux above the photospheric prediction.


Inspired by \citet{Xu2020}, we used the median and dispersion of $\chi$ to define a $3\sigma$ criterion. Instead of trimming the top and bottom $10\%$, we adopted the robust standard deviation $\sigma = 1.4826 \,\chg{\Delta_{\rm MAD}}$, with \chg{$\Delta_{\rm MAD}$} being the median absolute deviation; a relation that is satisfied when the population is Gaussian.

We find that the dispersion $\sigma_\chi$ of $\chi$ grows with the magnitude, an effect already observed by \citet{Madurga2024} who, to avoid the problems that would result from using a single $3\sigma_\chi$ threshold, defined two different magnitude bins: one for $G < 17$ and the other \chg{for $17 \leq G \leq 18.5$}.

\begin{figure*}
    \centering
    \includegraphics[width=\textwidth]{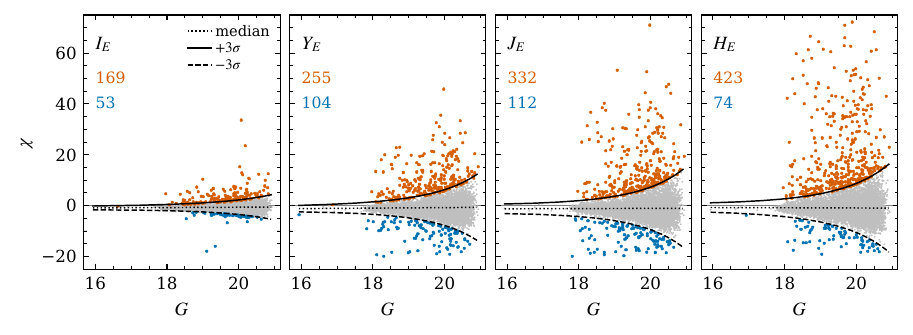}
    \caption{Flux-excess significance $\chi$ versus \textit{Gaia} \chg{$G$} magnitude in the four \chg{\Euclid} bands (\IE, \YE, \JE, \HE). The dotted line shows the fitted median of $\chi$, and the solid and dashed lines the median $\pm 3\sigma_\chi$, where $\sigma_\chi(G)$ is a quadratic fit in \chg{$\logten\sigma_\chi$} to the robust (MAD) dispersion in bins of 0.25. Orange and blue points mark sources with $z$-score $z > 3$ and $z < -3$, respectively; the numbers give their counts. About ten sources per NIR band with $\chi > 75$ lie outside the plotted range.}
    \label{fig:fitted_excesses_thresholds}
\end{figure*}

Since our WD sample extends down to the \textit{Gaia} limit of $G \approx 21$, over which $\sigma_\chi$ keeps increasing, a continuous parameterisation of $\sigma_\chi$ as a function of $G$ is more accurate. We therefore computed, for each band, the median of $\chi$, hereafter $\tilde{\chi}$, and $\sigma_\chi$ in $G$ bins of 0.25 between 18 and 21. Subsequently, we fitted \chg{$\logten\sigma_\chi$} and $\tilde{\chi}$ to a quadratic and a linear polynomial, respectively, for each band. We show the fitted curves in \chg{Fig.}~\ref{fig:fitted_excesses_thresholds} and their coefficients in \chg{Table~\ref{tab:excess_fits}}.

\begin{table}
\caption{Coefficients of the fits to the median and robust dispersion of the flux-excess significance $\chi$ as a function of $G$.}
\label{tab:excess_fits}
\centering
\small

\begin{tabular}{l c c c c c}
\toprule
Band & \multicolumn{1}{c}{$a_2$} & \multicolumn{1}{c}{$a_1$} & \multicolumn{1}{c}{$a_0$} & \multicolumn{1}{c}{$b_1$} & \multicolumn{1}{c}{$b_0$} \\
\midrule
\chg{$\IE$} & 0.02176 & $-0.6400$ & 4.0757 & 0.0700 & $-2.0435$ \\
\chg{$\YE$} & 0.02246 & $-0.6246$ & 3.8780 & 0.1022 & $-2.8412$ \\
\chg{$\JE$} & 0.02672 & $-0.8023$ & 5.8022 & 0.0401 & $-1.9138$ \\
\chg{$\HE$} & 0.02052 & $-0.5606$ & 3.5130 & $-0.0461$ & $-0.0233$ \\
\bottomrule
\end{tabular}
\tablefoot{The dispersion of the excess significance is modelled as $\chg{\logten}\sigma_\chi = a_2 G^2 + a_1 G + a_0$ and the median as $\tilde{\chi} = b_1 G + b_0$.}
\end{table}

Now we can compute the \chg{normalised} $\chi$ in each band $b$ by taking the $z$-score \chg{$z(b)= [\chi(b) - \tilde{\chi}(b)]/\sigma_\chi(b)$}, and we consider that a source has an excess in $b$ if \chg{$z(b) > 3$}.

A source is flagged as showing \chg{an} NIR excess if its \IE\ flux is consistent with the model \chg{[$|z(\IE)|<3$]} and it satisfies one of the following \chg{ conditions}: (i) \chg{$z(\JE) > 3$ and $z(\HE) > 3$}; or (ii) \chg{$z(\HE)>3$ with $z(\YE)<3$ and $z(\JE)<3$}. In total, 300 sources out of \chg{$11\,644$} \chg{(2.6\%)} with consistent \IE\ fluxes meet this criterion.

This should be regarded as an upper limit on the fraction of excesses detectable at this threshold, since noise fluctuations and residual contamination scatter sources above $z=3$. Estimating the intrinsic fraction would also require accounting for completeness, which is beyond the scope of this work. For reference, reversing the sign of the criterion ($z<-3$) yields 62 sources (0.5\%) with an apparent NIR deficit, indicating the size of the noise and systematics contribution.

We highlight that this fraction is \chg{considerably} lower than the \chg{approximately 6\%} typically \chg{reported} in the literature \citep[see\chg{,} e.g.\chg{,}][]{murillo2026}, a difference we attribute partly to the absence of mid-IR bands, but also to the angular resolution of \textit{Euclid}, which avoids the blends that likely inflate excess fractions derived from WISE.

\begin{figure*}
    \centering
    \includegraphics[width=\textwidth]{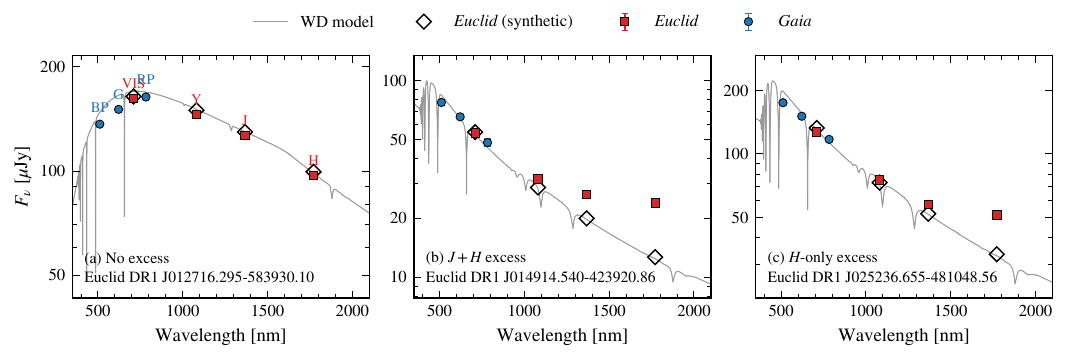}
    \caption{\chg{Example SEDs of three WDs in our sample}: no excess (left), excess in both $\JE$ and $\HE$ (centre), and excess only in $\HE$ (right).}
    \label{fig:wd_excess_examples}
\end{figure*}




\subsection{Sensitivity to atmospheric composition}

Most of our WDs lack spectral types, so we adopted H-rich models, which apply to most WDs: $77.5\%$ of the 40\,pc sample have H-dominated atmospheres \citep{Roberts2025}. Indeed, if we cross-match the \Euclid WD sample with the Montreal White Dwarf Database \citep[MWDD; ][]{Dufour2017}, we find, after removing uncertain spectral labels (those followed by a colon, ``:''): 3246 DAs ($84\%$), 263 DCs (featureless spectra; $7\%$), 259 DBs (He-rich; $7\%$), 70 DZs (metal-polluted; $2\%$), and 38 DQs (carbon-rich; $1\%$). Hence, the template model fitting is appropriate for \chg{approximately 84\%} of the sources.

To further quantify the impact of using H-rich models on non-DAs, we performed the same steps described in \chg{Sects.~\ref{sec:wd_model_fitting} and \ref{sec:nir_excesses}} but this time using the grids of 1D helium-rich (DB/DBA/DC) local thermodynamic equilibrium (LTE) spectra of \citet{Cukanovaite2021} and the $T_{\rm eff}$ and $\log{g}$ values computed by \citet{gf2021} assuming mixed atmospheres \chg{[$\logten(N_{\rm H}/N_{\rm He}) = -5$]} to obtain the model fluxes. As a result, we obtained 187 IR excesses, as opposed to the 300 computed with the H-rich models.

We note that 170 sources show excesses regardless of the assumed atmospheric composition, meaning that 130 sources that show excess using H-rich models do not show it with the He-rich models. Conversely, 17 IR excesses found with He-rich models are not present in the H-rich counterparts.

However, after checking the available spectral types, we find that DAs dominate in the latter subsets. Out of the 130 sources whose excess disappears with He-rich models, 27 have spectral types, and 26 of these are DAs. Of the 17 excesses that appear only with He-rich models, 8 of the 10 with spectral types are DAs. Therefore, we consider that the use of \citet{Tremblay2011} H-rich models does not have a catastrophic effect on the IR excess identification of He-rich WDs. \chg{In contrast,} \citet{murillo2026} \del{disagree with the previous claim. They} fitted He-rich WDs with the non-DA models of \citet{koester2010} and found IR excesses that would have gone undetected if DA models had been used. Overall\chg{,} this shows how model-dependent the identification of IR excesses is.

\subsection{Contamination by resolved neighbours}
\label{sec:photometric_contamination}

Nearby sources are a known source of spurious IR excesses in photometric
searches for cool companions to WDs \citep[e.g.\chg{,}][]{steele2011,murillo2026}.
In our sample, light from a resolved neighbouring source can enter the
\texttt{A-PHOT} aperture of the WD candidate and mimic the NIR excess expected
from an unresolved companion \citep{Merlin2019}. For this to happen, the
neighbour must be bright enough to supply the measured excess and close enough
for its light to enter the target aperture. We tested these two conditions for
all 300 candidates using the \Euclid MER catalogue \citep{Q1-TP004}.

For each target $i$ and band $b$, the flux that would need to be supplied by a
contaminating source is
\begin{equation}
\chg{E_i(b)=F_i^{\rm obs}(b)-F_i^{\rm mod}(b)}\;,
\label{eq:contamination_excess}
\end{equation}
where \chg{$F_i^{\rm obs}(b)$} is the corrected \texttt{A-PHOT} flux used for the
selection and \chg{$F_i^{\rm mod}(b)$} is the single-WD model flux. The uncertainty
in \chg{$E_i(b)$} corresponds to the denominator of Eq.~\eqref{fluxexcess}.
Thus, \chg{$E_i(b)$} is not the total target flux, but only the flux above the WD
model that a neighbouring source would have to reproduce. We tested only the
bands through which each source was selected: $\JE$ and $\HE$ for the 192
targets selected in both bands, and $\HE$ alone for the 108 $\HE$-only targets. A neighbour affecting only $\YE$ cannot explain the $\JE$ or $\HE$ excess
that defines the selection, so $\YE$ was not used in the contamination test.
The resulting classification therefore refers to the selecting bands and does
not imply that the $\YE$ measurement is free from neighbour light.

Considering only the targets tested in each band, the median excesses correspond to 41.0\% and 45.5\% of the observed flux in $\JE$ and $\HE$, respectively, while the smallest measured values are 12.4\% and 14.1\%. A fixed flux-ratio threshold would therefore be poorly suited to our sample: even a neighbour supplying $10\%$ of the target flux could not reproduce any candidate’s central excess. We therefore compared each neighbour directly with the measured excess of its target.

We searched the MER catalogue within $3''$ of every candidate and found
373 distinct target--neighbour pairs. For the candidate itself, we used
the \texttt{A-PHOT} flux from the NIR-excess selection, because this aperture
measurement can contain light from nearby sources. For each neighbour, we used
the NIR \texttt{T-PHOT} flux, which uses information from a
higher-resolution image to separate the flux of overlapping sources
\citep{Merlin2015,Merlin2016}, to estimate how much flux each
resolved neighbour could contribute to the \texttt{A-PHOT} excess.

For both target and neighbour measurements, we required finite uncertainties,
a zero spurious-source flag (\texttt{spurious\_flag = 0}), and band-specific and detection-quality flags of
0 or 2. The value 2 denotes a blended source\footnote{\url{https://euclid.esac.esa.int/dr/q1/dpdd/merdpd/merphotometrycookbook.html}}
and was retained because blends are directly relevant to the contamination
test. Positive evidence for contamination additionally required a positive
neighbour flux with $\mathrm{S/N}\geq5$.

The MER $2\times{\rm FWHM}$ aperture has a radius of one FWHM, corresponding to a median radius of 1\farcs3 for our sample \citep{Q1-TP004}.
A neighbour met the contamination criterion
only if it had reliable photometry, could reproduce the central measured excess
in at least one band through which the target was selected, lay inside the
target aperture, and did not share a valid MER deblend parent with the target.
A neighbour that failed this test was not automatically considered harmless;
a separate test was used to determine whether it could be ruled out.

To rule out a neighbour, we deliberately overestimated the amount of its light
that could contaminate the target. We assumed that its entire \texttt{T-PHOT}
flux entered the target aperture, irrespective of separation, and applied the
same aperture-to-total correction used for the target so that both quantities
were compared on the same flux scale. We then increased the possible neighbour
contribution by five standard deviations and reduced the target excess by one
standard deviation. A neighbour was ruled out only when its upper flux
estimate remained below the required excess in every band through which the
target was selected.

After considering all neighbours, we assigned each target one of three
classifications.
\begin{itemize}
    \item A target was classified as \chg{`contaminated'} if at least
    one neighbour had reliable photometry, lay inside the aperture, did not share
    a valid MER deblend parent with the target, and could supply the central
    measured excess.

    \item A target was classified as \chg{`clean'} if it had no resolved
    neighbour within $3''$, or if all its neighbours could be ruled out, both
    individually and with their upper contributions summed in each selecting band.
    A neighbour sharing a valid MER deblend parent with the target was never ruled
    out, even when its flux was too faint.

    \item All remaining targets were classified as
    \chg{`ambiguous'}: no contaminating neighbour had been identified, but the
    catalogue information was insufficient to establish a clean result.
\end{itemize}

We identify 107 clean candidates (35.7\%) and 193 ambiguous candidates
(64.3\%), while none meets the contamination criterion
(Table~\ref{tab:contamination}). Out of the 107 clean targets, 71 have no separate MER neighbour within $3''$.
The remaining 36 contain one or more resolved neighbours, but the upper flux
estimates of their neighbours, individually and summed, remain below the
required excess in the relevant bands.

\begin{table}
\caption{Resolved-neighbour contamination classification for the 300
NIR-excess candidates. The subrows give the full composition of each
category. Ambiguous targets are listed by the reasons that prevented a clean
classification; ``only'' means that no other reason was recorded.}
\label{tab:contamination}
\centering
\small
\renewcommand{\arraystretch}{1.2}
\begin{adjustbox}{max width=\columnwidth}
\begin{tabular}{lr}
\toprule
Category or subset & Number \\
\midrule
Clean & 107 \\
\quad No resolved neighbour within $3''$ & 71 \\
\quad All resolved neighbours ruled out & 36 \\
Ambiguous & 193 \\
\quad Shared deblend parent only & 163 \\
\quad Shared parent and neighbour upper flux sufficient & 8 \\
\quad Shared parent and indeterminate neighbour photometry & 8 \\
\quad Neighbour upper flux sufficient only & 4 \\
\quad Indeterminate neighbour photometry only & 9 \\
\quad Summed neighbour upper flux not ruled out & 1 \\
Contaminated & 0 \\
\midrule
Total & 300 \\
\bottomrule
\end{tabular}
\end{adjustbox}
\end{table}

\begin{figure}[ht]
\centering
\includegraphics[width=\columnwidth]{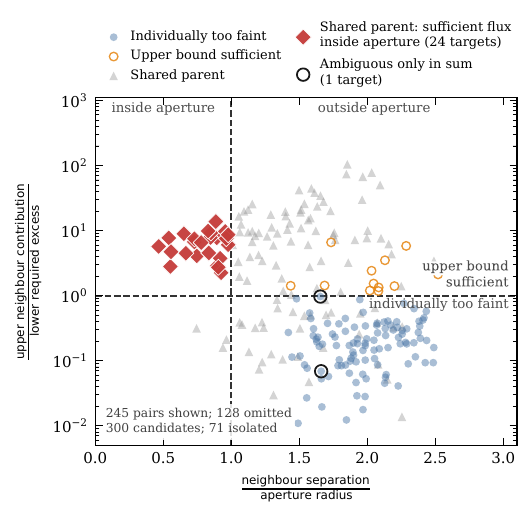}
\caption{Resolved-neighbour flux test for the 245 of the 373
target--neighbour pairs with a usable $\JE$ or $\HE$ comparison; each point is
one pair. The horizontal axis is the separation in units of the target
aperture radius, and the vertical axis is the neighbour's conservative upper
contribution divided by the $1\sigma$ lower required excess, in the selecting
band where this ratio is largest. Points below the horizontal dashed line are individually too faint to explain
the excess. Inside the aperture, all neighbours with reliable measured flux
sufficient to reproduce the excess share a MER deblend parent with the target
(red diamonds, 24 systems). The two outlined points are the neighbours of one target: both lie outside the
aperture and are individually too faint, but their summed upper contributions
prevent a clean classification.}
\label{fig:contamination_screen}
\end{figure}

\begin{figure*}[ht]
\centering
\includegraphics[width=\textwidth]
{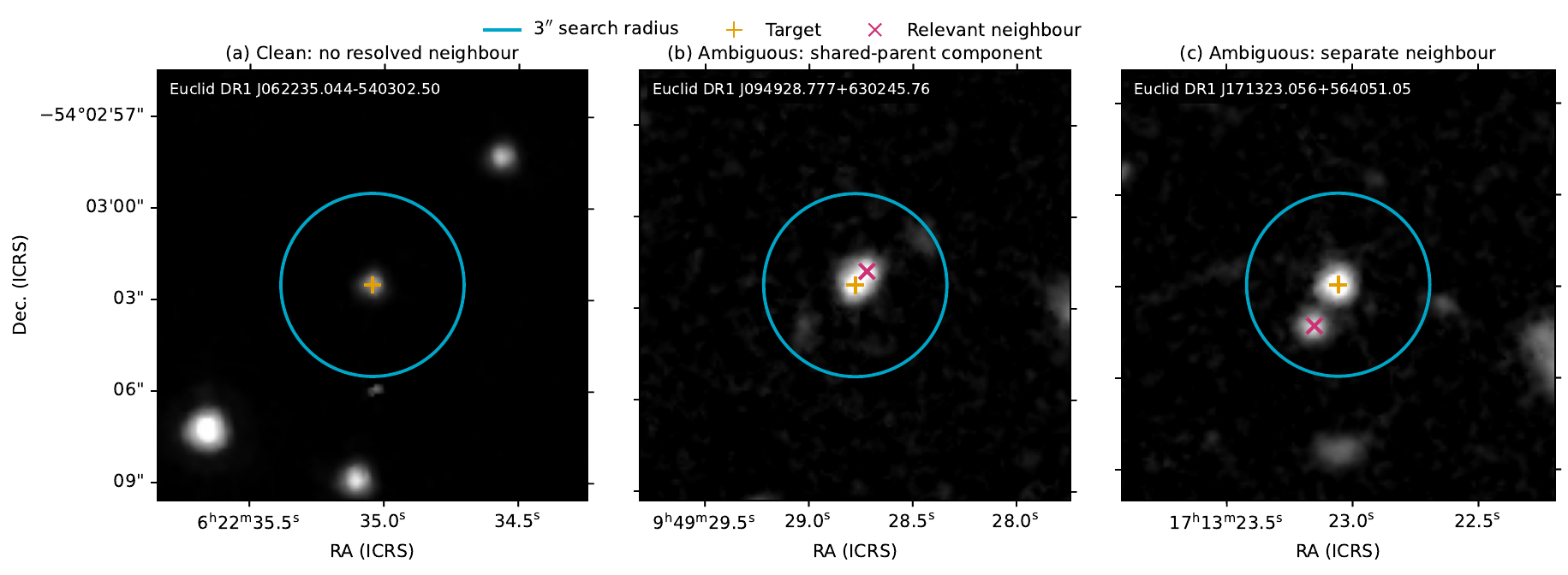}
\caption{Representative \Euclid NISP $\HE$ cutouts illustrating three
local source configurations. Panel (a) shows an isolated target with no
resolved MER neighbour within $3''$. In panel (b), the target and nearby
component share a MER deblend parent and are therefore treated conservatively
rather than as independent evidence of contamination. In panel (c), the nearby
source does not share a MER deblend parent with the target and is therefore
treated as a distinct resolved neighbour.}
\label{fig:three_images}
\end{figure*}

Most ambiguous targets (179 of 193) contain a component that shares the
target's MER deblend parent. Among these, 24 come closest to satisfying the
positive contamination criterion. Each contains a component inside the target aperture
with reliable flux sufficient to reproduce the central measured excess in a
selecting $\JE$ or $\HE$ band
(Fig.~\ref{fig:contamination_screen}). However, the target and neighbouring
component share the same valid MER \texttt{PARENT\_ID}, indicating that they
are deblended components associated with the same parent
source\chg{.}\footnote{\url{https://euclid.esac.esa.int/dr/q1/dpdd/merdpd/dpcards/mer_finalcatalog.html}}
We therefore do not treat these components as independent evidence that a distinct neighbouring source causes the excess, and retain all 24 systems as
ambiguous. No neighbour with a separate MER deblend parent satisfies the flux,
quality, and aperture conditions simultaneously. Representative isolated, shared-parent, and separate-neighbour configurations
are shown in Fig.~\ref{fig:three_images}.

We also tested whether several neighbours that were individually too faint
could reproduce the excess together. Summing the reliable central fluxes of
separately deblended neighbours inside the target aperture produces no
additional contamination case. Summing the conservative upper contributions
of all neighbours within $3''$, separately in each selecting band, affects one
of the four otherwise clean targets with more than one neighbour: its summed
upper bound reaches 1.04 times the lower required excess, so we classify it as
ambiguous. Both of its neighbours lie outside the aperture, so it does not
meet the contamination criterion.

In the individual-neighbour test, before the summed
check, the number of clean targets is 108 for the $1\sigma$ lower bound, the
central measured excess, and the $2\sigma$ lower bound alike, but increases to
223 when the search radius is reduced from $3''$ to 1\farcs5. We retain $3''$
as the baseline because we do not model how much of a neighbour's PSF enters
the target aperture as a function of separation. Before the summed check, removing the shared-parent restriction at the $3''$ search radius would reclassify 24 of the 192 ambiguous targets as contaminated and 16 as clean. The remaining 152 would stay ambiguous.

We retain all 300 NIR-excess candidates in the catalogue and provide
the NIR-excess selection and contamination classification as flags
(Table~\ref{tab:cds_columns}). The following spectroscopic and variability analyses focus on the 107 clean candidates.

\section{Spectroscopic characterisation}\label{sec:nisp_spectra}
 
\subsection{Template fitting}\label{sec:template_fitting}
 
We retrieved the \Euclid NISP red-grism spectra for the 107 sources with a clean excess. The red grism covers 1200--1900\,nm at $R\approx450$, enough to detect the H$_2$O and CH$_4$ molecular bands, as demonstrated by \citet{Dominguez2025}. Four sources do not have NISP spectra, leaving 103 objects for the subsequent analysis. We treated all pixels with \texttt{QUALITY<0.7} as artefacts and removed them. We also removed pixels with bad bits set in the \texttt{MASK} extension (odd values or values $\geq 64$).
 
\begin{figure}[h]
    \centering
    \includegraphics[width=\hsize]{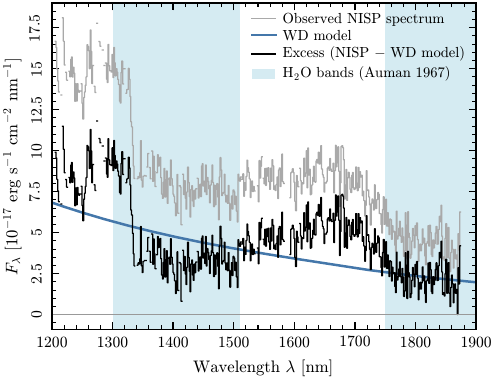}
    \caption{Example of the WD subtraction for Euclid\,DR1\,J034847.177$-$423122.35. The observed NISP red-grism spectrum (grey), the best-fitting WD model (blue), and the residual excess (black) are shown at native resolution. Shaded regions mark the H$_2$O absorption bands at 1300--1510 and 1750--2050\,nm \citep{Auman1967}.}
    \label{fig:nisp_spectra_vs_wdmodel}
\end{figure}

For each object, we subtracted the best-fitting WD model flux (Sect.~\ref{sec:wd_model_fitting}) from the observed NISP spectrum. The residual, $E_\lambda = F_\lambda^{\rm NISP} - F_\lambda^{\rm WD}$, is the spectrum of the unresolved companion. Figure~\ref{fig:nisp_spectra_vs_wdmodel} shows an example of this subtraction for Euclid\,DR1\,J034847.177$-$423122.35, which also has strong H$_2$O bands.
 
To characterise the companions, we fitted the subtracted spectra with the M, L, and T dwarf spectral standards of the SpeX Prism Library Analysis Toolkit \citep[SPLAT;][]{Burgasser2017}, using $\chi^2$ minimisation in the \texttt{splat} Python package.\footnote{\url{https://github.com/aburgasser/splat/}} Before the comparison, we degraded each excess spectrum to $R\approx120$ (about 50 points), comparable to the resolution of the SPLAT standards, and restricted it to 1225--1845\,nm to avoid the edges. Despite the faintness of our sample (median $\HE\approx20.2$), 58 of the 103 spectra were suitable for template fitting; we discarded the other 45 for low S/N or incomplete coverage.

\subsection{Overview of the fits}\label{sec:fit_overview}
 
Visual inspection divides the 58 fits into two groups. For 18 sources, the SPLAT standards reproduce the excess satisfactorily, and we consider their spectral types confirmed (tier~1). The remaining 40 spectra mostly show clear, structured features, but either no field standard reproduces their shape within the uncertainties or the wavelength coverage is too incomplete to constrain the type, so we give only a candidate classification (tier~2). Appendix~\ref{appendix_C} shows the fits of both groups. The confirmed companions range from M1 to T1 (Fig.~\ref{fig:fits_tier1}).

We additionally compared the absolute magnitude of each companion, $M_{H_\mathrm{E}}$, computed from the excess flux and the \textit{Gaia} distance, with that expected for its best-fitting spectral type from the field-age $H_\mathrm{E}$ relation of \citet{sanghi2024}, which has a root-mean-square scatter of 0.44\,mag. This relation is valid for types M6--T9, which covers 13 of the 18 confirmed companions. For 11 of them, the observed and expected values agree within $2\sigma$, where $\sigma$ combines the photometric and distance uncertainties with the scatter of the relation. The other two, the M8--L0 Euclid DR1 J043819.361$-$470629.74 and the L3--L7 Euclid DR1 J040749.253$-$524436.44, are fainter than expected for their best-fitting types but agree with the late end of their spectral ranges, which we therefore favour.

 
We note that SPLAT standards are spectra of isolated field dwarfs, whereas irradiated companions can show large, wavelength-dependent day--night differences \citep{Casewell2015}. For SDSS\,J1411+2009, \citet{Lew2021} found variations of $(360\pm70)\%$ in the H$_2$O band depth and $(37\pm2)\%$ in the $J$ band. The phase-averaged spectrum of a strongly irradiated BD can therefore depart from every isolated-BD template, so sources that no template reproduces may be candidates for the most strongly irradiated companions.

\subsection{L and T dwarf companions}\label{sec:LT_dwarfs}
 
\begin{figure*}[t]
  \centering
  \includegraphics[width=\textwidth]{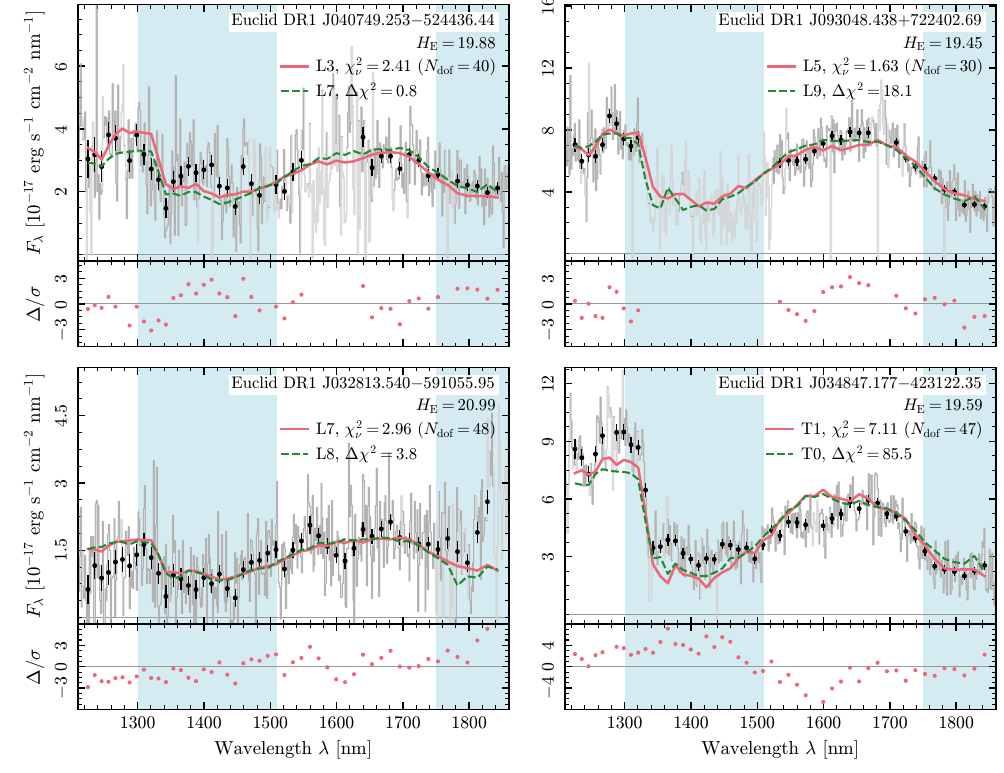}

  \caption{Template fits of the four BD companions among the tier-1 sources (Sect.~\ref{sec:LT_dwarfs}). Each panel shows the NISP spectrum after subtracting
    the WD model, smoothed to $R\approx 120$ (black points), together with the
    best-fitting (red, solid) and next-best (green, dashed) SpeX spectral standards \citep{Burgasser2017}. The legend gives the reduced $\chi^2_\nu$ of the best fit and the
    $\Delta\chi^2$ of the next-best one. Lower panels show the residuals of the best
    fit in units of $\sigma$. Light-blue regions mark the H$_2$O absorption bands.}
  \label{fig:best_fits}
\end{figure*}
 
Four of the confirmed companions are BDs (Fig.~\ref{fig:best_fits}): three have best-fitting types later than L4, and the fourth is an L3--L7 dwarf whose absolute magnitude favours the late end of its range (Sect.~\ref{sec:fit_overview}).

\begin{itemize}

\item Euclid DR1 J040749.253$-$524436.44 shows a clear \JE peak, H$_2$O trough, and \HE bump. Part of the \HE peak is masked, so L3 and L7 templates fit equally well ($\Delta\chi^2 = 0.8$, $N_{\rm dof}= 40$). Its absolute magnitude, $M_{H_\mathrm{E}} = 15.1\pm 0.2$, is too faint for an L3 dwarf but instead agrees with the late end of this range, so we consider the companion a BD.

\item Euclid\,DR1\,J093048.438+722402.69 is best fitted by an L5 template ($\chi^2_{\nu}=1.63$, $N_{\rm dof}=30$). Its H$_2$O band at $1.3\,\mu$m has few good pixels, but its absorption is clearly visible at native resolution and in the $1.75\,\mu$m band.
 
\item Euclid\,DR1\,J032813.540$-$591055.95 has one of the faintest excesses in the sample, which makes its spectrum noisy. Despite this, its H$_2$O trough and \HE peak are reproduced by an L7 template ($\chi^2_{\nu}=2.96$, $N_{\rm dof} = 48$), with L8 as a close alternative ($\Delta\chi^2 = 3.8$).
 
\item Euclid\,DR1\,J034847.177$-$423122.35 (see also Fig. \ref{fig:nisp_spectra_vs_wdmodel}) shows the deepest H$_2$O absorption among the 18 confirmed systems and a pronounced \HE-band peak. It is fitted as a T1 dwarf, strongly preferred over the next-best T0 template ($\Delta\chi^2 = 85.5$, $N_{\rm dof} = 47$).
 
\end{itemize}
 
The T1 dwarf adds to the three known unresolved WD--T systems, WD\,0837+185 \citep{Casewell2012}, SDSS\,J1411+2009 \citep{Littlefair2014, Casewell2018}, and ZTF\,J0038+2030 \citep{vanRoestel2021, Broski2026}, all of which were discovered through radial velocities or eclipses rather than through their NIR excess.
 
Several tier-2 spectra also show strong H$_2$O bands and L/T-like shapes (Fig.~\ref{fig:fits_tier2}). The two clearest T-dwarf candidates are Euclid DR1 J173548.343+541424.00 and Euclid DR1 J043937.463$-$191518.65. The first is best fitted by a T4 template, but with large residuals ($\chi^2_\nu = 15.4$, $N_{\rm dof} = 40$), and its companion would be $5\sigma$ brighter than a T4 dwarf at the distance of the WD, consistent instead with an early-L dwarf. The second shows a narrow $H_\mathrm{E}$ peak and an absolute magnitude consistent with a late-T dwarf (best fit T8), but also an unexplained feature at 1440--1470\,nm. A red point source about 22 times brighter in \HE lies $3\farcs9$ away, beyond our $3''$ screening radius, and shares its MER deblend parent. This source most likely contaminates the slitless spectrum. The L/T-like spectra of Euclid DR1 J155200.104+780611.51 (Sect.~\ref{sec:variability_results}) and Euclid DR1 J023937.102$-$580042.95 (Sect.~\ref{sec:paschen}) are too noisy or affected by artefacts for a reliable fit. NIR spectroscopy that resolves these systems at higher S/N is needed to firmly establish their nature.

\subsection{Systems with Paschen emission}\label{sec:paschen}
 
\begin{figure}
    \centering
    \includegraphics[width=\hsize]{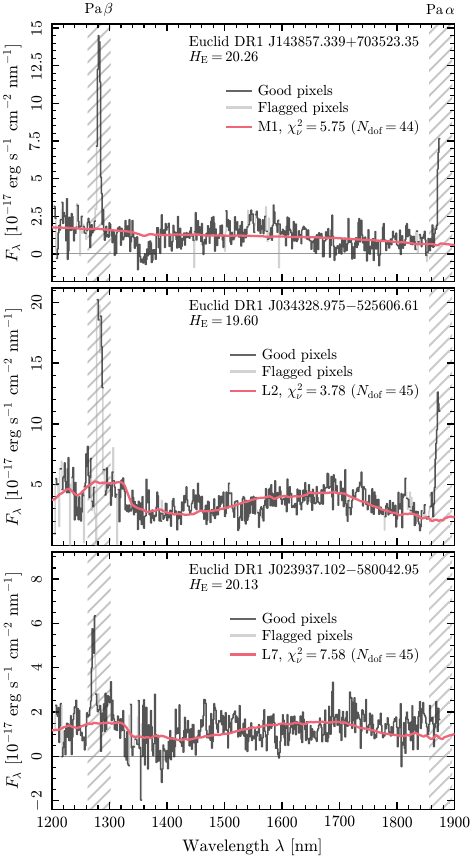}
    \caption{Subtracted NISP spectra of the three WDs with Pa\,$\beta$ emission, shown at native resolution. The template fits were performed at $R\approx120$ with the Pa\,$\beta$ and Pa\,$\alpha$ regions (hatched) masked.}
    \label{fig:pab_emitters}
\end{figure}

Three sources show strong Pa\,$\beta$ emission, and partly Pa\,$\alpha$ (Fig.~\ref{fig:pab_emitters}), which are compatible with the accretion signatures expected in CVs. We fitted their spectra at $R\approx120$ with both Paschen lines masked.
 
\begin{itemize}
\item Euclid\,DR1\,J143857.339+703523.35 has no previous references in the literature, and without time-domain photometry we cannot classify it firmly as a CV. However, its \textit{Gaia} low-resolution spectrum also shows a tentative H\,$\alpha$ emission line, which makes it a strong new CV candidate with a tentative M1 donor ($\chi^2_\nu = 5.75$, $N_\mathrm{dof} = 44$; Fig.~\ref{fig:pab_emitters}).

\item Euclid\,DR1\,J034328.975$-$525606.61 was identified as a CV candidate by \citet{Hodgkin2021} and later confirmed as a period-bouncer CV by \citet{Munoz2026}. Period bouncers are CVs that have evolved past the minimum orbital period, after the donor has been stripped to a very low mass and the orbit has begun to widen. \citet{Munoz2026} used photometric SED fitting to constrain the donor to a T dwarf. Our template fit (tier~1) instead favours an L2--L3 donor ($\chi^2_\nu = 3.78$, $N_\mathrm{dof} = 45$; Fig.~\ref{fig:pab_emitters}), which places the companion close to the hydrogen-burning limit.

\item Euclid\,DR1\,J023937.102$-$580042.95 is a new CV candidate with a tentative, tier~2, substellar donor: L7 with the Paschen lines masked, L5 without ($\chi^2_{\nu}=7.58$ with $N_{\rm dof}=45$).
\end{itemize}

\section{Variability analysis}
\label{sec:variability}

A companion that explains the NIR excess need not be in a close orbit. At the median distance of approximately $300\,\mathrm{pc}$ for the \Euclid WD sample (Table~\ref{tab:representativeness}), $1''$ corresponds to approximately $300\,\mathrm{au}$. By comparison, orbital periods from a few hours to ten days imply separations of at most a few tens of solar radii for typical WD and low-mass companion masses. A companion can therefore remain unresolved while being too widely separated to produce detectable orbital variability.

Time-domain photometry provides an independent way of identifying close systems through orbital variability. Relevant signatures include eclipses, irradiation-driven modulation, and ellipsoidal modulation caused by tidal distortion \citep{hilditch2001}. Irradiation-driven orbital modulation is observed in close WD--BD binaries such as \chg{WD\,0137$-$349} and GD~1400AB, with amplitudes that can vary substantially between systems \citep[e.g.\chg{,}][]{Casewell2015,Casewell2024,Amaro2025}. We therefore searched the available light curves of our candidates for periodic variability.

\subsection{Light-curve data}

We selected all 107 candidates classified as clean in Sect.~\ref{sec:photometric_contamination}, regardless of their spectroscopic results. The 193 ambiguous candidates were not included in the search. 

No single survey has the cadence, precision, and sky coverage needed for every target. We therefore used data from 12 archives. The ground-based data came from the All-Sky Automated Survey for Supernovae (ASAS-SN; \citealt{kochanek2017,hart2023}), the Asteroid Terrestrial-impact Last Alert System (ATLAS; \citealt{tonry2018}), BlackGEM \citep{groot2024}, the Catalina Real-Time Transient Survey (CRTS; \citealt{drake2009}), the Javalambre Variability Survey (J-VAR; \citealt{ederoclite2025}), the NOIRLab Source Catalog (NSC; \citealt{nidever2021}), the Optical Gravitational Lensing Experiment (OGLE; \citealt{udalski2015}), SkyMapper \citep{onken2024}, and the Zwicky Transient Facility (ZTF; \citealt{bellm2019,masci2019}). The space-based data came from \chg{\textit{Gaia}} \citep{gaiadr3}, K2 \citep{howell2014}, and the Transiting Exoplanet Survey Satellite (TESS; \citealt{ricker2015}).

The surveys use different photometric units, time systems, and quality flags. We retrieved and standardised the light curves with \texttt{lcquery} \citep{lcquery}\chg{,}\footnote{\url{https://github.com/andrijazupic/lcquery}}  a Python package for multi-survey time-domain light-curve retrieval and standardisation.  We matched targets by \chg{\textit{Gaia}} DR3 source identifier where supported. Otherwise, we used coordinate searches or forced photometry. For coordinate-based retrieval, we propagated the \chg{\textit{Gaia}} positions to adopted survey reference epochs using the available proper motions. The package applied the recommended survey-specific quality cuts, converted the photometry to flux densities in $\mu\mathrm{Jy}$, and converted the times to barycentric Julian dates on the barycentric dynamical time scale (BJD$_{\rm TDB}$).

\subsection{Period search}
The search covered $10\,\mathrm{min}$ to $10\,\mathrm{d}$, encompassing the hour-to-day periods of PCEBs \citep{nebot2011}. We searched every band with at least ten valid measurements, using all retained epochs. To preserve eclipses, we clipped bright outliers beyond six robust standard deviations but did not apply the same cut to faint measurements. We searched each band separately because the variability amplitude, sampling, and precision can differ between bands. 

We used four complementary period-finding methods to search for smooth modulation and narrow eclipses. Lomb--Scargle \citep[LS;][]{lomb1976,scargle1982} fits a sinusoid at each trial frequency and is suited to approximately sinusoidal variability. Box-fitting least squares \citep[BLS;][]{kovacs2002} searches for periodic box-shaped dips in flux and is suited to narrow eclipses. Phase dispersion minimisation \citep[PDM;][]{stellingwerf1978} minimises the scatter within phase bins without assuming a light-curve shape. Finally, we used a phase-binned adaptation of the kernel-regression approach implemented in \texttt{FINKER} \citep{stoppa2024}, using a Gaussian kernel to search for more complex periodic variability.

We prioritised signals when at least two period estimates from different methods, bands, or surveys agreed within 0.2\%, allowing agreement at the same period and at the $2\times$ and $\tfrac{1}{2}\times$ harmonics. These harmonic alternatives account for systems with two similar brightness dips or peaks per orbit, for which a method can recover half the orbital period \citep{Graham2013}. Over a long observing baseline, small differences between period estimates can lead to large differences in the predicted number of cycles. After allowing for the factors of two, we therefore checked whether the predicted cycle counts differed by less than half a cycle over the longer baseline. We gave higher priority to agreements that met this condition and were supported by different bands or surveys.

We also prioritised single strong peaks with a nominal LS false-alarm probability below $10^{-3}$ \citep{baluev2008}, or a signal detection efficiency (SDE) of at least 15 for BLS, 10 for PDM, or 9 for the kernel statistic. We calculated SDE by subtracting a median-based periodogram baseline and dividing by the standard deviation of the residual periodogram. These thresholds ranked results for inspection; SDE values are descriptive scores, not false-alarm probabilities.

For every band and method, we inspected the complete light curves folded at the best period. We also inspected the highest-ranked distinct alternative whose SDE was at least 90\% of that of the best peak. We retained a signal when at least one well-sampled band showed coherent variability. We rejected incoherent folds and signals explained by daily, monthly, or other cadence-related aliases. We rejected marginal signals confined to one band when other data of comparable quality showed no corresponding variation. A clear signal could still be retained when the other bands lacked the cadence or precision to test it.

\subsection{Screening for variable neighbours}

We also checked whether a variable neighbour could explain each retained signal. The MER classification in Sect.~\ref{sec:photometric_contamination} applies to the \Euclid NIR photometry. The time-domain surveys have different angular resolutions and extraction apertures, so we assessed possible blends separately for each survey. We searched SIMBAD \citep{Wenger2000} and the International Variable Star Index \citep[VSX;][]{watson2006} within $2'$ of each target, accounting for proper motion. We assessed potential contaminants within a survey-dependent screening radius, $r_{\rm cont}$: $3''$ for OGLE, NSC, and BlackGEM; $3\farcs5$ for \chg{\textit{Gaia}}; $4''$ for ZTF and J-VAR; $6''$ for SkyMapper and CRTS; $8''$ for ATLAS; $17''$ for K2; $25''$ for ASAS-SN; and $60''$ for TESS. We chose these radii conservatively to allow for each survey's PSF, photometric aperture, or extraction scale. They are screening limits, not exact aperture boundaries.

We compared catalogue periods with the detected period and with one-third, one-half, twice, and three times that period, using a tolerance of 0.2\%. We reviewed sources within $2''$ as possible counterparts of the target. We flagged a distinct variable as a possible contaminant if its period matched and it lay within the applicable $r_{\rm cont}$. This check cannot exclude contamination by an uncatalogued variable neighbour.

\subsection{Results}\label{sec:variability_results}

\begin{figure*}[ht]
\centering
\includegraphics[width=0.95\textwidth]{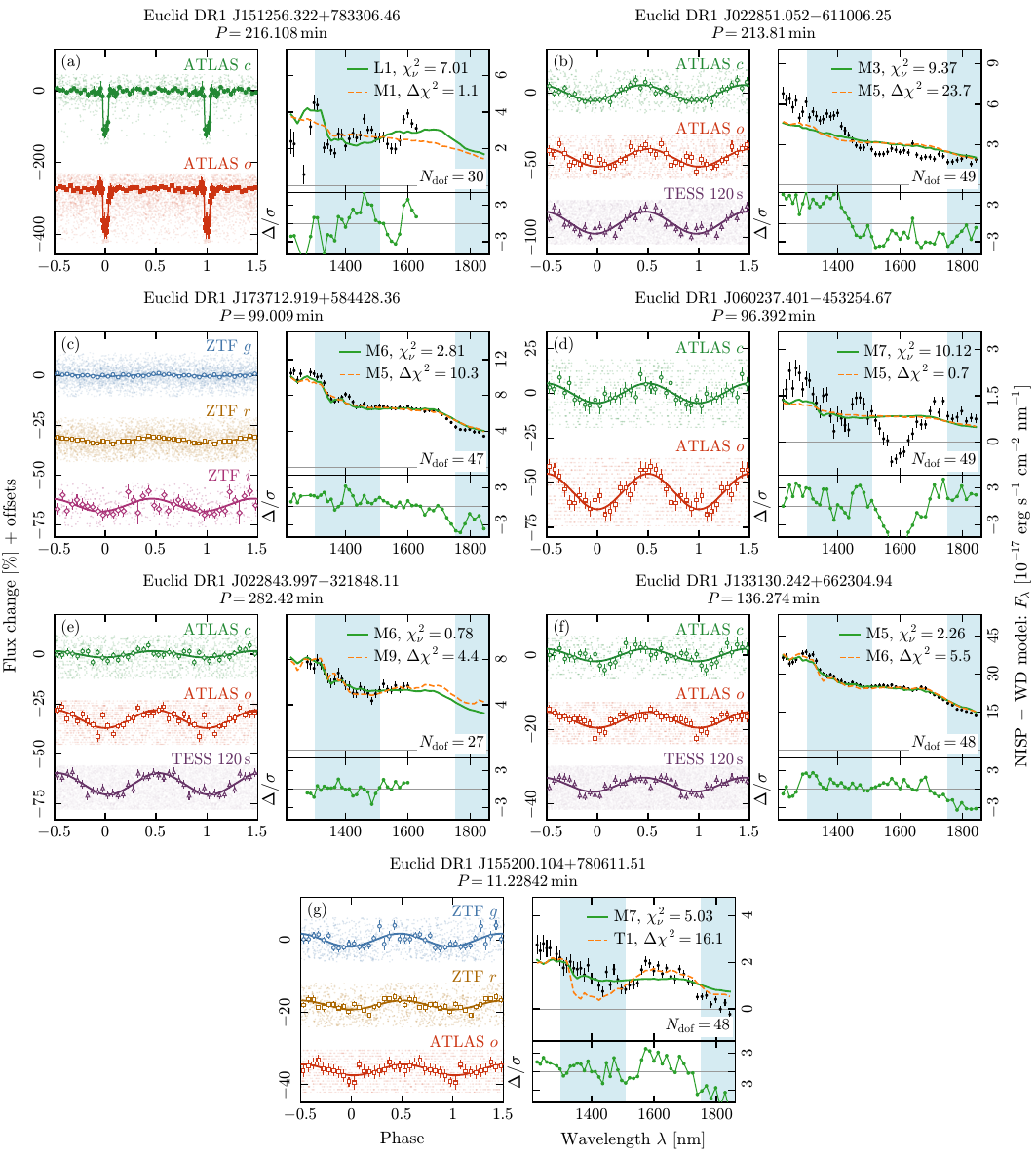}
\caption{Light curves and NISP spectra of six periodic detections and one rapid-variability candidate (g). Each source is shown as a pair: folded light curves on the left and the WD-subtracted spectrum on the right. The light curves span two cycles. Faint points show individual measurements, and symbols with error bars show phase-binned means. Curves show sinusoidal fits, except in (a), where lines connect the eclipse bins. Flux changes are percentages of the median flux. Within each source, bands are offset vertically and share a flux scale and phase zero. The spectral fits, residuals, and shading are presented as described in Fig.~\ref{fig:best_fits}.}
\label{fig:variability_example}
\end{figure*}

Among the 107 clean candidates analysed, we find coherent periodic variability in six systems and a rapid-variability candidate in a seventh (Table~\ref{tab:periods}, Fig.~\ref{fig:variability_example}). The SIMBAD and VSX queries did not identify any of these seven targets as previously known variables. They also returned no catalogued variable neighbour with a matching period within the applicable screening radii. The six detections are marked by the \texttt{Periodic} flag in the catalogue described in Table~\ref{tab:cds_columns}.

\begin{table*}[ht]
\caption{Periodic signals among the 107 clean candidates.}
\label{tab:periods}
\centering
\small
\setlength{\tabcolsep}{4pt}
\renewcommand{\arraystretch}{1.2}
\begin{adjustbox}{max width=\textwidth}
\begin{tabular}{llcccccc}
\toprule
Euclid DR1 designation & \chg{\textit{Gaia}} DR3 & SpT
& $P$~[min] & Signature & Band & Amplitude~[$\%$] & Amplitude S/N\\
\midrule
J151256.322+783306.46 & \texttt{1708381550460691072} & L1: & 216.108 & eclipse & ATLAS $c$ & $108\pm5$ & 21.2 \\
& & & & & ATLAS $o$ & $105\pm5$ & 19.1 \\
\midrule
J022851.052$-$611006.25 & \texttt{4725496535738551296} & M3: & 213.81 & quasi-sinusoidal & ATLAS $c$ & $5.5\pm0.8$ & 7.0 \\
& & & & & ATLAS $o$ & $6.1\pm0.9$ & 6.7 \\
& & & & & TESS & $7.6\pm1.0$ & 7.8 \\
\midrule
J173712.919+584428.36 & \texttt{1434569654399191936} & M6 & 99.009 & quasi-sinusoidal & ZTF $g$ & $0.38\pm0.19$ & 2.0 \\
& & & & & ZTF $r$ & $1.40\pm0.23$ & 6.0 \\
& & & & & ZTF $i$ & $3.2\pm0.8$ & 3.9 \\
\midrule
J060237.401$-$453254.67 & \texttt{5567675610291663360} & L:
& 96.392 & quasi-sinusoidal & ATLAS $c$ & $5.5\pm1.1$ & 5.1 \\
& & & & & ATLAS $o$ & $9.9\pm1.1$ & 9.3 \\
\midrule
J022843.997$-$321848.11 & \texttt{5063589901649049856} & M6 & 282.42 & quasi-sinusoidal & ATLAS $c$ & $1.6\pm0.6$ & 2.9 \\
& & & & & ATLAS $o$ & $4.5\pm0.6$ & 7.3 \\
& & & & & TESS & $5.4\pm1.0$ & 5.2 \\
\midrule
J133130.242+662304.94 & \texttt{1672332721713582976} & M5 & 136.274 & quasi-sinusoidal & ATLAS $c$ & $1.7\pm0.5$ & 3.7 \\
& & & & & ATLAS $o$ & $2.11\pm0.34$ & 6.2 \\
& & & & & TESS & $1.92\pm0.30$ & 6.4 \\
\midrule
J155200.104+780611.51 & \texttt{1707782007385659904} & T1: & $11.22842^{\dagger}$ & quasi-sinusoidal & ZTF $g$ & $1.8\pm0.4$ & 5.0 \\
& & & & & ZTF $r$ & $1.25\pm0.29$ & 4.3 \\
& & & & & ATLAS $o$ & $1.5\pm0.4$ & 3.5 \\
\bottomrule
\end{tabular}
\end{adjustbox}
\tablefoot{The adopted photometric period is $P$. Amplitudes are eclipse-core depths or fitted sinusoidal semi-amplitudes, expressed as percentages of the median retained flux. Eclipse depths above 100\% result from slightly negative measured core fluxes. Uncertainties are estimated standard errors, allowing for residual scatter and correlations within observing nights. Amplitude S/N is $A/\sigma_A$ at fixed $P$, calculated before rounding. The dagger marks the rapid-variability candidate: fits use $P=11.22842\,\mathrm{min}$, but sampling aliases remain unresolved. We also include the adopted spectral type (`SpT'). A colon marks uncertain types.}
\end{table*}

\begin{itemize}
\item Euclid\,DR1\,J151256.322+783306.46 shows a deep eclipse in both ATLAS bands, with near-zero flux at minimum and no detectable secondary eclipse (Fig.~\ref{fig:variability_example}a, left panel). The odd- and even-cycle eclipses have compatible depths. If the orbital period were $2P$, these would be eclipses of different stars. The two stars cannot each supply nearly all of the out-of-eclipse light, so we adopt $P=216.108\,\mathrm{min}$ as the orbital period. The NISP spectrum is noisy and incomplete for this source (Fig.~\ref{fig:variability_example}a, right panel), and thus the best-fitting spectral type is degenerate between L1 and M1 ($\chi^2_{\nu} = 7.01$ and $\Delta\chi^2=1.1$ with $N_{\rm dof}=30$), needing follow-up spectroscopy to confirm the true nature of this companion.

\item Euclid\,DR1\,J022851.052$-$611006.25 shows quasi-sinusoidal modulation with an adopted period of $P=213.81$\,min and with compatible amplitudes and phases in ATLAS $c$, ATLAS $o$, and TESS (Fig.~\ref{fig:variability_example}b). In this case, the NISP spectrum favours an M-dwarf, but the fit shows large residuals with correlated structure ($\chi^2_{\nu}=9.37$ with $N_{\rm dof}=49$), indicating that the template is most likely not the best one.

\item Euclid\,DR1\,J173712.919$+$584428.36 shows lower-amplitude modulation, clearest in ZTF $r$ and $i$ (Fig.~\ref{fig:variability_example}c) with $P=99$\,min. Its semi-amplitude increases towards redder bands at compatible phases (Table~\ref{tab:periods}). This is consistent with variability from a cool companion, because constant WD light can dilute the modulation more strongly in the blue. The NISP spectrum also supports an M6 companion (Sect.~\ref{sec:nisp_spectra}).

\item Euclid\,DR1\,J060237.401$-$453254.67 shows quasi-sinusoidal modulation in both ATLAS bands (Fig.~\ref{fig:variability_example}d) with the adopted period being $P=96.39$\,min. The bands agree in phase, with a larger fitted semi-amplitude in $o$ than in $c$. The NISP spectrum shows an L/T-dwarf shape in the \JE region, while the deep V-shape valley at 1.6\,$\mu$m (arguably due to slitless spectra issues because of the measured negative flux) prevents a good fit and favours an M7 type with large residuals ($\chi^2_{\nu}=10.12$ with $N_{\rm dof}=49$) instead.

\item Euclid\,DR1\,J022843.997$-$321848.11 shows quasi-sinusoidal modulation, clearest in ATLAS $o$ and TESS, with compatible amplitudes and phases (Fig.~\ref{fig:variability_example}e). TESS favours the adopted period ($P=282.42$\,min) over the tested daily aliases. ATLAS $c$ and CRTS provide weaker support. In this case, the NISP spectrum is best fitted by a late-M template (M6--M9; $\chi^2_{\nu}=0.78$ and $\Delta\chi^2=4.4$ with $N_{\rm dof} = 27$), but without \HE coverage we consider the type tentative (tier~2).

\item Euclid\,DR1\,J133130.242+662304.94 shows quasi-sinusoidal modulation at $P=136.274\,\mathrm{min}$, with compatible amplitudes and phases in ATLAS $c$, ATLAS $o$, and TESS (Fig.~\ref{fig:variability_example}f). The NISP spectrum shows an M5 companion ($\chi^2_{\nu} = 2.26$ with $N_{\rm dof} = 48$), the likely source of this variability.
\end{itemize}

For the five smooth detections with hour-scale periods, the variability mechanism remains uncertain. If the modulation is orbital, irradiation would imply an orbital period equal to the photometric period, whereas ellipsoidal modulation would imply twice that period.

The eclipse provides independent evidence for a close binary. The five smooth modulations with hour-scale periods are consistent with close companions. We did not use non-detections to reject hidden companions. Even a close binary can escape detection at an unfavourable inclination or with insufficient cadence or photometric precision. We searched only clean candidates, and the sensitivity varied between targets. The recovered fraction therefore does not measure the intrinsic variable fraction of all 300 NIR-excess sources.

One additional source stands out from the six systems discussed above because its variability occurs on a much shorter timescale. Euclid\,DR1\,J155200.104+780611.51 shows a candidate $P=11.23\,\mathrm{min}$ modulation in ZTF $g$ and $r$, with tentative support from ATLAS $o$ at a compatible phase (Fig.~\ref{fig:variability_example}g). No convincing longer-period explanation emerged from the alias and harmonic checks, although the exact period remains uncertain. Such a short period is incompatible with a detached WD--BD orbit, since the companion would overflow its Roche lobe \citep{Rappaport2021}.
Instead, the $674\,\mathrm{s}$ timescale and fitted WD temperature of $T_{\rm eff}\simeq11\,400\,\mathrm{K}$ \citep{gf2021} place the WD in the ZZ Ceti strip. ZZ Ceti stars are hydrogen-atmosphere WDs that undergo non-radial pulsations with periods of a few minutes as they cross the WD instability strip \citep{Bognar2020}. The tentative increase in amplitude from ZTF $r$ to $g$ is also consistent with variability originating from the WD. Its WD-subtracted NISP spectrum nevertheless shows a T-like shape, with T1 as the second-best-fitting template and the most plausible substellar interpretation. We therefore treat this source as a rapid-variability candidate, most likely a ZZ Ceti, rather than as an orbital-period detection.

\section{Conclusions}
\label{sec:conclusions}

In this work we have carried out a search for unresolved cool companions to \chg{WDs} through \chg{NIR} excesses in Euclid DR1. We assessed contamination by resolved neighbours and characterised
the clean candidates through NISP spectroscopy. We also
searched their archival light curves for periodic variability. We summarise our main conclusions below.

\begin{itemize}

\item By cross-matching the \chg{\textit{Gaia}} WD catalogue of \citet{gf2021} with the Euclid DR1 MER catalogue, we built a sample of 13\,122 WDs over \chg{approximately} 1900\,deg$^2$ of the Euclid Wide Survey. Apart from the lower extinction expected from its high-Galactic-latitude footprint and a modest \chg{(6\%)} shift towards larger distances, the sample traces the parent \chg{\textit{Gaia}} WD population without \chg{appreciable} bias in its intrinsic properties.

\item We used WD models anchored to \textit{Gaia} photometry to predict the photospheric fluxes in the \Euclid $I_\mathrm{E}$, $Y_\mathrm{E}$, $J_\mathrm{E}$, and $H_\mathrm{E}$ bands and compared them with the \Euclid \texttt{A-PHOT} photometry using a magnitude-dependent parameterisation of the flux-excess significance. We find 300 WDs with NIR excesses at the $3\sigma$ level, corresponding to a NIR excess fraction of $2.6\%$. This is between two and three times lower than the $\approx 6\%$ commonly reported in the literature, which we attribute to the absence of mid-IR bands in our selection and to the angular resolution of \Euclid, which avoids the blends that likely inflate WISE-based excess fractions.

\item The resolved-neighbour screen classifies 107 (35.7\%) NIR-excess candidates as clean and 193 (64.3\%) as ambiguous; none meets our contamination criterion. We retain all 300 candidates in the catalogue with the classification recorded as a flag. The spectroscopic and variability analyses focus on the 107 clean candidates.

\item Of the 103 clean candidates with NISP spectra, 58 have sufficient S/N and spectral coverage for template fitting, and 18 pass our visual inspection as confirmed cool companions, with spectral types from M1 to T1. Four of them are brown dwarfs: an L5 (Euclid DR1 J093048.438+722402.69), an L7--L8 (Euclid DR1 J032813.540$-$591055.95), an L3--L7 whose absolute magnitude favours the late end (Euclid DR1 J040749.253$-$524436.44), and a T1 dwarf (Euclid DR1 J034847.177$-$423122.35). Several other candidates were identified and await confirmation.

\item Three systems show Pa\,$\beta$ emission, evidencing accretion: a known period bouncer with an L2--L3 donor and two new CV candidates with tentative M1 and L5--L7 donors.

\item A search of archival light curves of the 107 clean candidates identifies coherent periodic variability in six systems. One is an eclipsing binary with an orbital period of $216.1\,\mathrm{min}$. The other five show smooth modulation at photometric periods between $96.4$ and $282.4\,\mathrm{min}$; their orbital periods may equal these values or twice them. A seventh source shows a tentative signal near $11.23\,\mathrm{min}$, without an assigned orbital interpretation. No catalogued variable neighbour with a matching period was found within the adopted screening radii.

\end{itemize}

\Euclid thus opens the faint end of the WD companion population to systematic study. Higher-S/N NIR spectroscopy of the remaining clean and ambiguous candidates will be needed to establish the nature of their excesses, and further time-domain observations will determine the mechanisms and orbital periods of the smooth variables. Future \Euclid data releases, covering the full $\approx 14\,000$\,deg$^2$ of the Wide Survey, will increase this sample by a factor of about seven and enable a completeness-corrected determination of the occurrence rate of low-mass and substellar companions to WDs.

%

\begin{acknowledgements}
\AckEC
\chg{\AckDRone}
XPC acknowledges financial support from the Spanish National Programme for the Promotion of Talent and its Employability grant PRE2022-104959 cofunded by the European Social Fund. AZ acknowledges financial support from the Spanish National Programme for the Promotion of Talent and its Employability grant PREP2024-002594 funded by ESF. XPC, AZ, MM, and CD acknowledge support from the Spanish Ministry of Science MCIN / AEI / 10.13039 / 501100011033 and the EU FEDER collaboration through the coordinated grant PID2024-157964OB-C22; from the Xunta de Galicia and the European Union (FEDER Galicia 2021-2027 Program) Ref. ED431B 2024/21, ED431B 2024/02, and CITIC ED431G 2023/01. M.F.T. acknowledges funding from the Agencia Estatal de Investigación (AEI) of the
Ministerio de Ciencia, Innovación y Universidades (MCIU) and the European Social
Fund Plus (ESF+) under grant JDC2024-055310-I. This work has made use of data from the European Space Agency (ESA) Gaia mission and processed by the Gaia Data Processing and Analysis Consortium (DPAC). Funding for the DPAC has been provided by national institutions, in particular, the institutions participating in the Gaia Multilateral Agreement.
\end{acknowledgements}

%
%

\bibliography{Euclid, DR1}
%
%

%

\begin{appendix}

\section{\texorpdfstring{\chg{ADQL} query in the \chg{Euclid Science} Archive}{ADQL query in the Euclid Science Archive}}\label{sec:adql_query}

\begin{verbatim}
SELECT *
FROM catalogue.mer_catalogue_wide_survey
WHERE vis_det = 1
AND flux_vis_2fwhm_aper > 0
AND flux_y_2fwhm_aper > 0
AND flux_j_2fwhm_aper > 0
AND flux_h_2fwhm_aper > 0
AND mer_catalogue_wide_survey.det_quality_flag < 3
AND mer_catalogue_wide_survey.gaia_id IS NOT NULL
\end{verbatim}

\section{Catalogue description}
\label{app:catalogue}

Table~\ref{tab:cds_columns} describes the columns of the catalogue of the 13\,122 \textit{Euclid} WDs available at the Centre de Données astronomiques de Strasbourg (CDS). Columns marked with a subscript $b$ are given for each of the four \textit{Euclid} bands, $b \in \{\IE, \YE, \JE, \HE\}$. Binary flags take the value 1 if the condition is met and 0 otherwise within their applicable samples; fields that do not apply to a given source are left empty. The contamination columns are only filled in for the 300 NIR-excess candidates. The \texttt{Periodic} flag is populated only for the 107 clean candidates searched for variability: 1 denotes a retained periodic signal and 0 denotes no retained signal. Entries are left empty for sources not included in the variability search.

\begin{table*}
\caption{Description of the columns of the \textit{Euclid} WD catalogue available at the CDS.}
\label{tab:cds_columns}
\centering
\begin{tabular}{l l l}
\hline\hline
Column & Unit & Description \\
\hline
\multicolumn{3}{l}{\textit{Identification}} \\
\texttt{EuclidID} & -- & Euclid DR1 MER \texttt{object\_id} \\
\texttt{GaiaDR3} & -- & \chg{\textit{Gaia}} DR3 \texttt{source\_id} \\
\hline
\multicolumn{3}{l}{\textit{Gaia astrometry and photometry}} \\
\texttt{RAdeg}, \texttt{DEdeg} & deg & MER right ascension and declination (ICRS) \\
\texttt{Gmag}, \texttt{BPmag}, \texttt{RPmag} & mag & \chg{\textit{Gaia}} DR3 photometry \\
\texttt{plx}, \texttt{e\_plx} & mas & Parallax and error corrected following \citet{Lindegren2021} and \citet{MaizApellaniz2022}  \\
\texttt{pmRA}, \texttt{pmDE} & mas\,yr$^{-1}$ & Proper motions corrected following \citet{Cantat-Gaudin2021} \\
\texttt{dist50}, \texttt{dist16}, \texttt{dist84} & pc & Distance \citep{Weiler2025} and its 16th and 84th percentiles \\
\hline
\multicolumn{3}{l}{\textit{WD parameters \citep[assuming H-rich; from][]{gf2021}}} \\
\texttt{PWD} & -- & Probability of being a WD\\
\texttt{Teff}, \texttt{e\_Teff} & K & Effective temperature \\
\texttt{logg}, \texttt{e\_logg} & -- & Surface gravity \\
\texttt{Mass}, \texttt{e\_Mass} & $M_\odot$ & Mass \\
\texttt{AV} & mag & Mean $V$-band extinction \\
\hline
\multicolumn{3}{l}{\textit{Euclid photometry and NIR excess}} \\
\texttt{F\_b}, \texttt{e\_F\_b} & $\mu$Jy & Corrected \chg{\texttt{A-PHOT}} total flux and uncertainty (Sect.~\ref{sec:euclid_photometry}) \\
\texttt{Fmod\_b} & $\mu$Jy & Synthetic flux of the H-rich WD model (Sect.~\ref{sec:wd_model_fitting}) \\
\texttt{alpha} & -- & Fitted normalisation \chg{$(R_{\rm WD}/d)^2$} \\
\texttt{chi\_b} & -- & Flux-excess significance (Eq.~\ref{fluxexcess}) \\
\texttt{z\_b} & -- & Normalised significance, \chg{$z(b) = [\chi(b) - \tilde{\chi}(b)]/\sigma_\chi(b)$} \\
\hline
\multicolumn{3}{l}{\textit{Selection flags}} \\
\texttt{TeffSamp} & -- & \chg{Has $T_{\rm eff}$ and $\log g$ in \citet{gf2021} and $\mathrm{S/N}>5$ in all bands} \\
\texttt{IEcons} & -- & \IE\ flux consistent with the model (\chg{$|z(\IE)|<3$}) \\
\texttt{NIRexc} & -- & Selected as NIR excess (Sect.~\ref{sec:nir_excesses}) \\
\texttt{ExcCrit} & -- & Selection criterion: \texttt{JH} ($\JE$ and $\HE$) or \texttt{H} ($\HE$ only) \\
\texttt{NIRdef} & -- & NIR deficit (sign-reversed criterion) \\
\hline
\multicolumn{3}{l}{\textit{Contamination by resolved neighbours (Sect.~\ref{sec:photometric_contamination})}} \\
\texttt{Contam} & -- & Classification: \texttt{clean} or \texttt{ambiguous} \\
\texttt{ContamCode} & -- & Reason for the classification (subcategories in Table~\ref{tab:contamination}) \\
\texttt{Nneigh} & -- & Number of resolved MER neighbours within 3\arcsec \\
\texttt{SharedParent} & -- & At least one neighbour shares the target's MER \texttt{PARENT\_ID} \\
\hline
\multicolumn{3}{l}{\textit{Ancillary information}} \\
\texttt{SpType} & -- & Spectral type from the MWDD if available \citep{Dufour2017} \\
\texttt{SpecFit} & -- & Adopted spectral template when available (Sect.~\ref{sec:nisp_spectra}) \\
\texttt{Periodic} & -- & Periodic signal retained (Sect.~\ref{sec:variability}) \\
\hline
\end{tabular}
\end{table*}

\section{Template fits}\label{appendix_C}
\subsection{Tier 1: good fits}
Figure~\ref{fig:fits_tier1} shows the 18 tier-1 spectra (Sect. \ref{sec:fit_overview}), including the four also shown in Fig.~\ref{fig:best_fits}. 

\begin{figure*}
  \centering
  \includegraphics[width=0.33\textwidth]{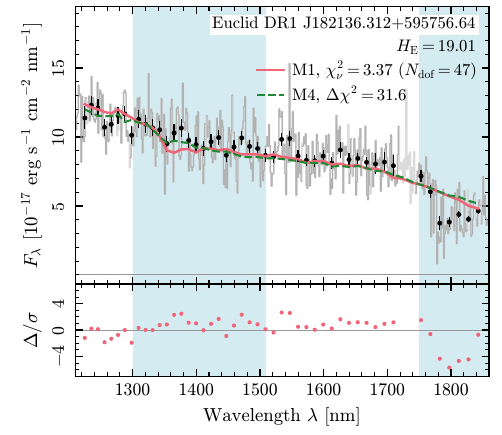}\hfill
  \includegraphics[width=0.33\textwidth]{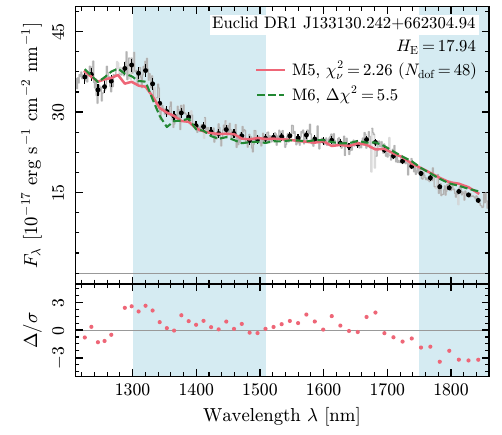}\hfill
  \includegraphics[width=0.33\textwidth]{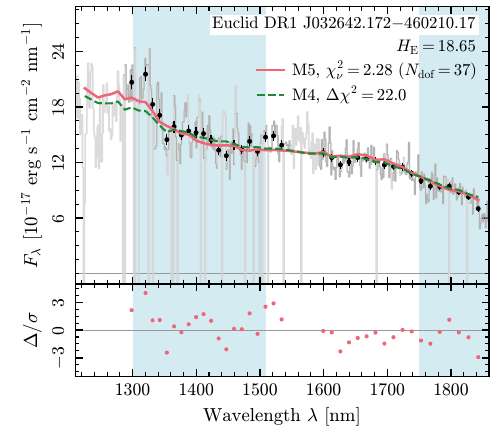}\\[2pt]
  \includegraphics[width=0.33\textwidth]{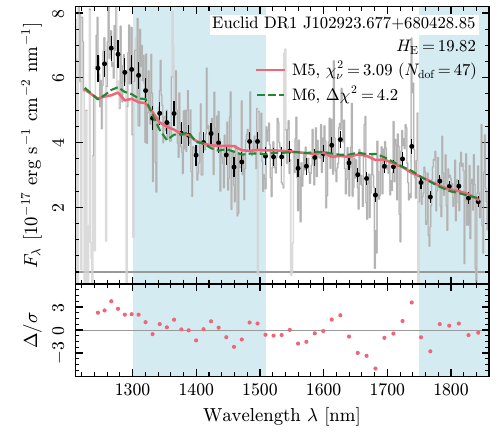}\hfill
  \includegraphics[width=0.33\textwidth]{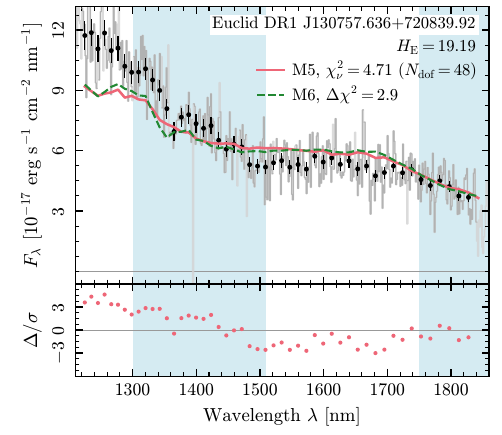}\hfill
  \includegraphics[width=0.33\textwidth]{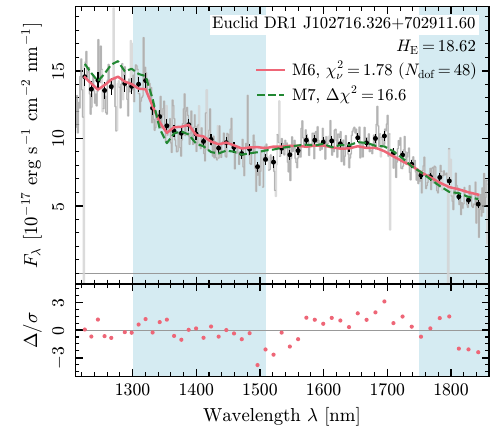}\\[2pt]
  \includegraphics[width=0.33\textwidth]{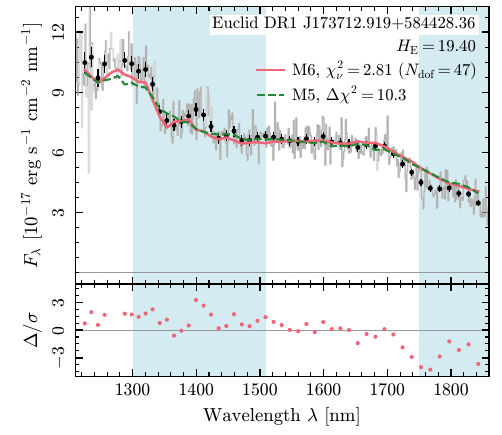}\hfill
  \includegraphics[width=0.33\textwidth]{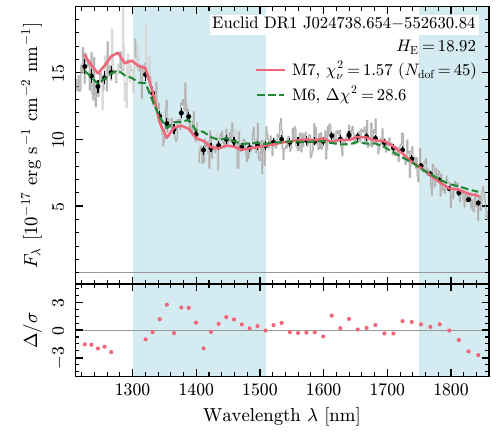}\hfill
  \includegraphics[width=0.33\textwidth]{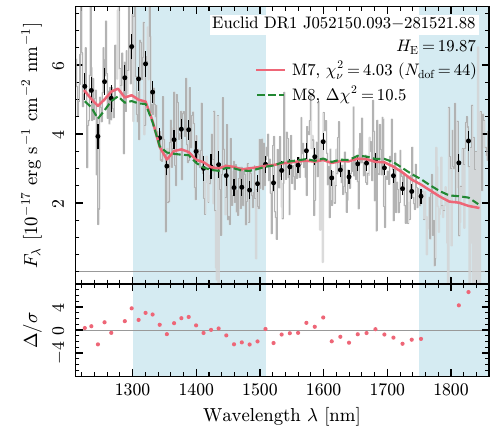}\\[2pt]
  \includegraphics[width=0.33\textwidth]{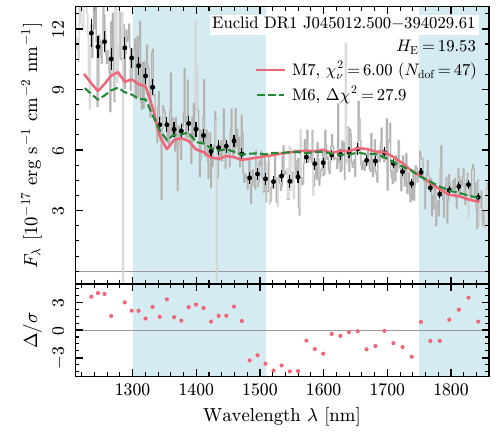}\hfill
  \includegraphics[width=0.33\textwidth]{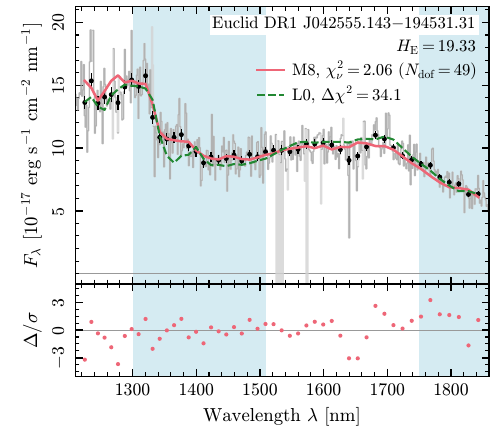}\hfill
  \includegraphics[width=0.33\textwidth]{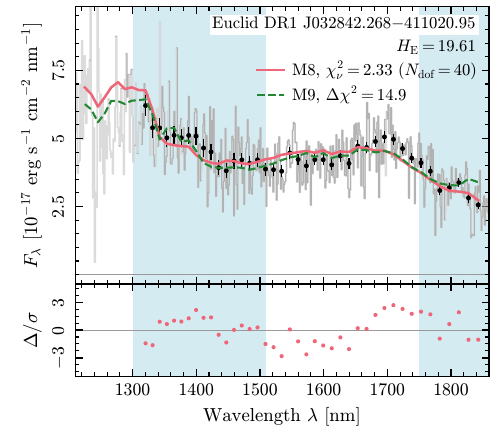}
  \caption{Infrared excess (NISP spectrum minus the WD model) of the tier 1 sources, sorted by best-fitting
  spectral type from hottest to coolest. Black points: excess rebinned to $R\approx120$; dark grey: native NISP
  resolution, good pixels only; light grey: native resolution including masked pixels. Solid red and dashed
  green lines: best-fitting and next-best SpeX spectral standards (labels give $\chi^2_\nu$ and $\Delta\chi^2$).
  Shaded regions: H$_2$O bands \citep{Auman1967}. Lower panels: residuals of the best fit in units of $\sigma$.}
  \label{fig:fits_tier1}
\end{figure*}
\begin{figure*}
  \addtocounter{figure}{-1}
  \renewcommand{\theHfigure}{\thefigure-cont}
  \centering
  \includegraphics[width=0.33\textwidth]{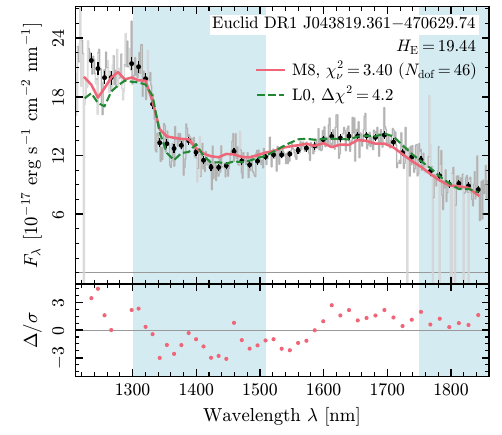}\hfill
  \includegraphics[width=0.33\textwidth]{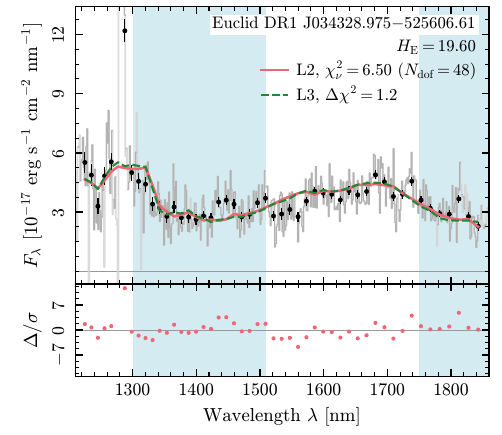}\hfill
   \includegraphics[width=0.33\textwidth]{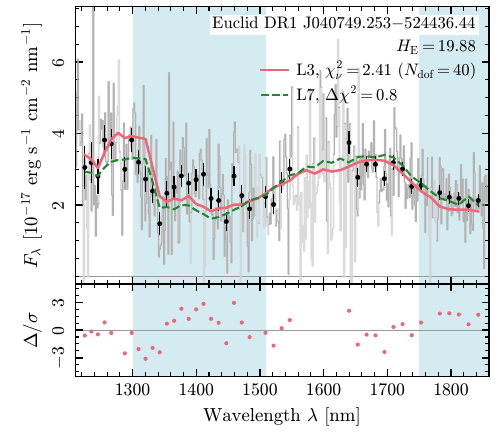}\\[2pt]
  \includegraphics[width=0.33\textwidth]{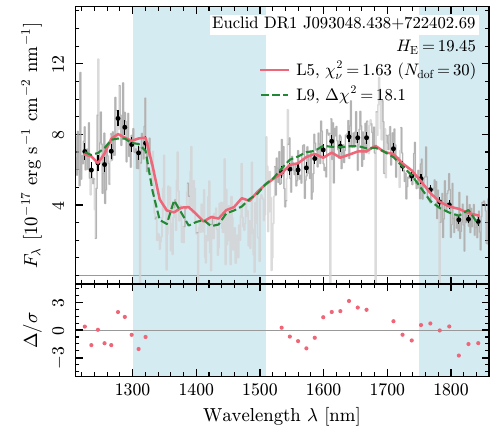}\hfill
  \includegraphics[width=0.33\textwidth]{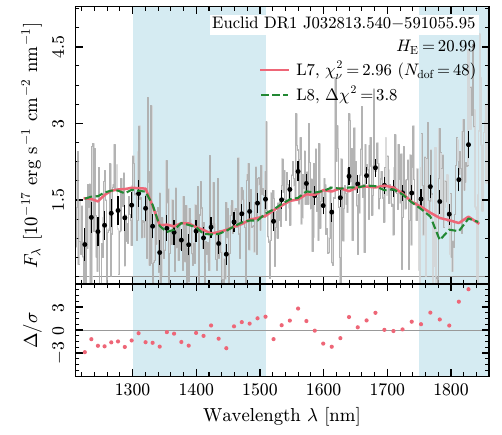}\hfill
  \includegraphics[width=0.33\textwidth]{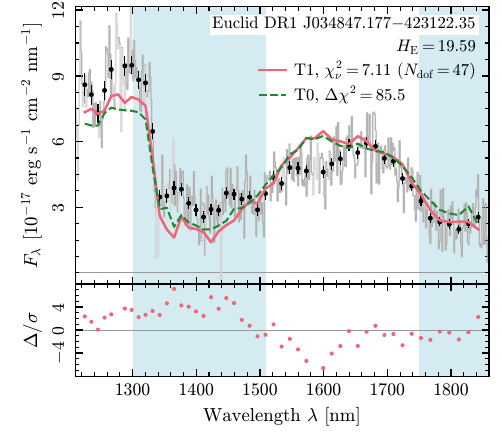}
  \caption{Continued.}
\end{figure*}

\subsection{Tier 2: candidate fits}

Figure~\ref{fig:fits_tier2} shows the 40 tier-2 spectra. The fits of the remaining 45 spectra, with low S/N or incomplete coverage, will be made available as supplementary material.

\begin{figure*}
  \centering
  \includegraphics[width=0.33\textwidth]{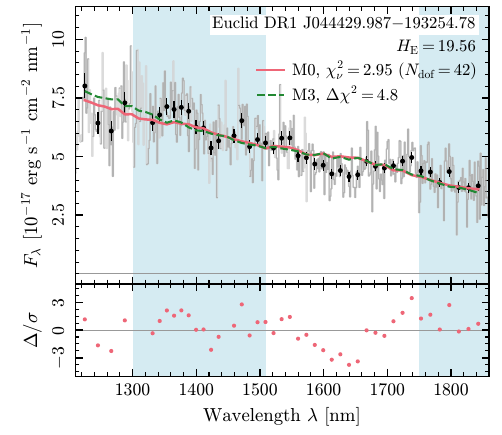}\hfill
  \includegraphics[width=0.33\textwidth]{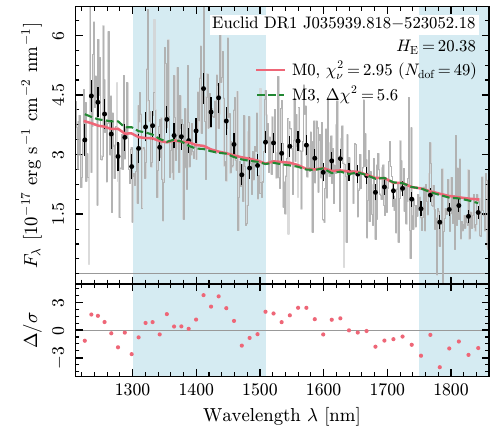}\hfill
  \includegraphics[width=0.33\textwidth]{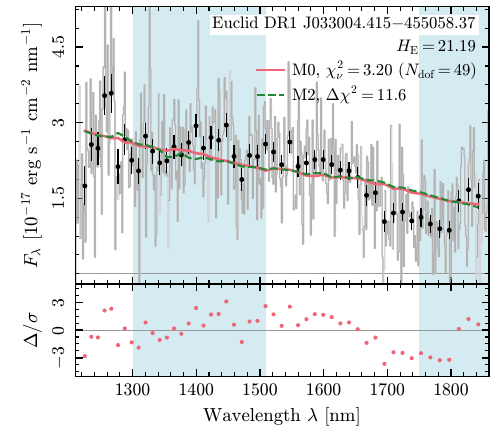}\\[2pt]
  \includegraphics[width=0.33\textwidth]{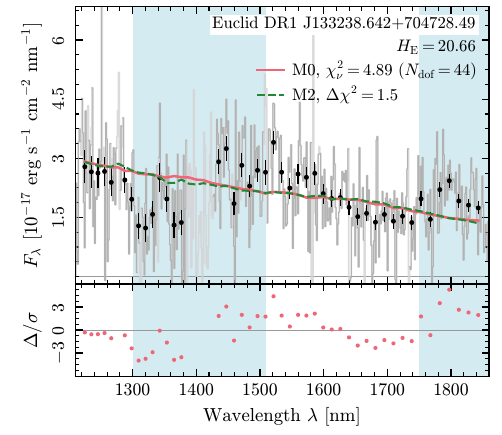}\hfill
  \includegraphics[width=0.33\textwidth]{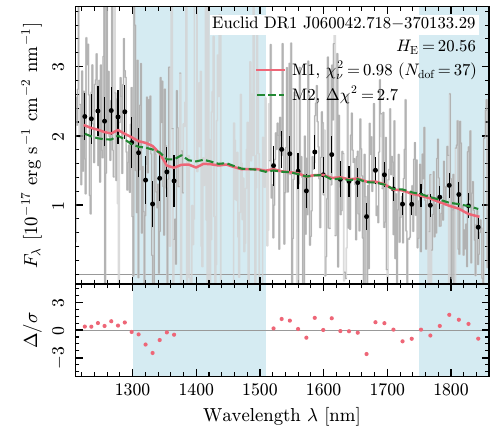}\hfill
  \includegraphics[width=0.33\textwidth]{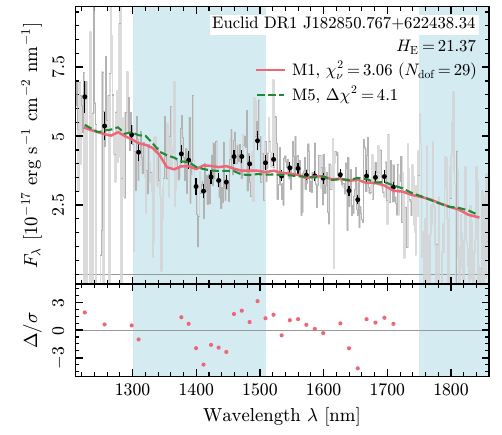}\\[2pt]
  \includegraphics[width=0.33\textwidth]{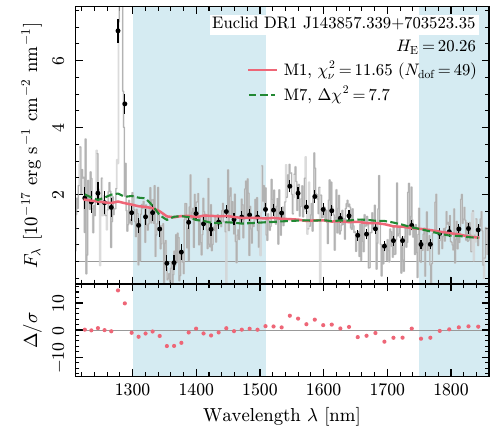}\hfill
  \includegraphics[width=0.33\textwidth]{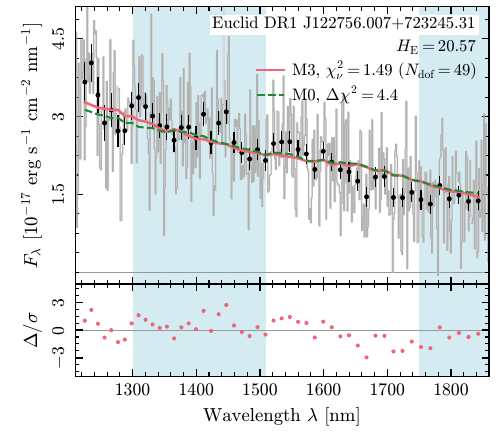}\hfill
  \includegraphics[width=0.33\textwidth]{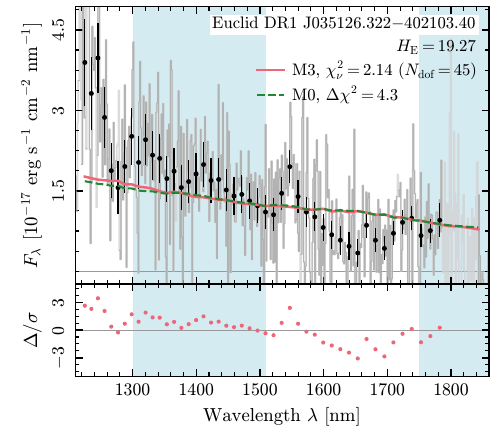}\\[2pt]
  \includegraphics[width=0.33\textwidth]{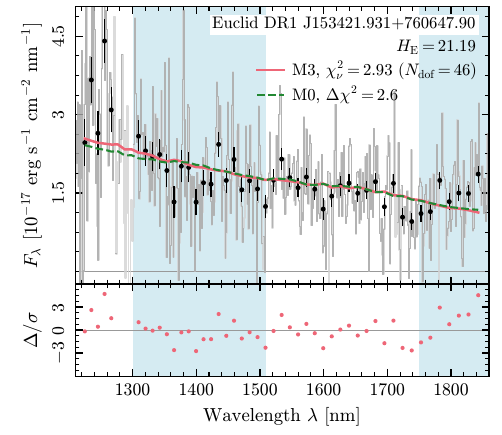}\hfill
  \includegraphics[width=0.33\textwidth]{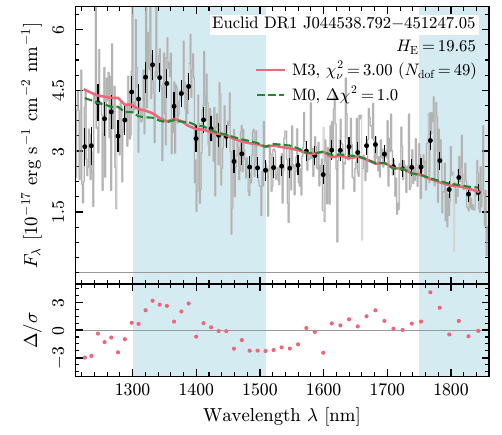}\hfill
  \includegraphics[width=0.33\textwidth]{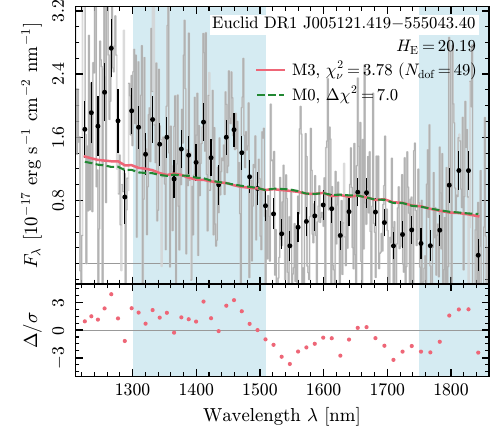}
  \caption{Infrared excess (NISP spectrum minus the WD model) of the tier 2 sources, sorted by best-fitting
  spectral type from hottest to coolest. Black points: excess rebinned to $R\approx120$; dark grey: native NISP
  resolution, good pixels only; light grey: native resolution including masked pixels. Solid red and dashed
  green lines: best-fitting and next-best SpeX spectral standards (labels give $\chi^2_\nu$ and $\Delta\chi^2$).
  Shaded regions: H$_2$O bands \citep{Auman1967}. Lower panels: residuals of the best fit in units of $\sigma$.}
  \label{fig:fits_tier2}
\end{figure*}
\begin{figure*}
  \addtocounter{figure}{-1}
  \renewcommand{\theHfigure}{\thefigure-cont}
  \centering
  \includegraphics[width=0.33\textwidth]{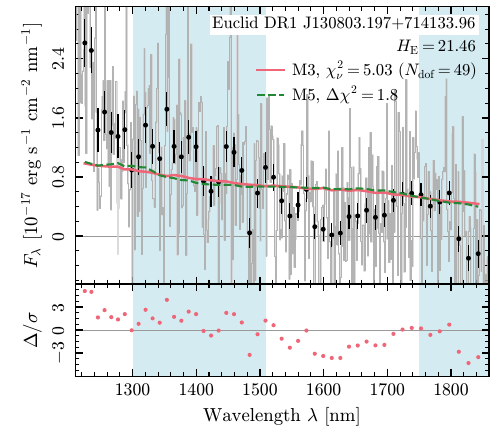}\hfill
  \includegraphics[width=0.33\textwidth]{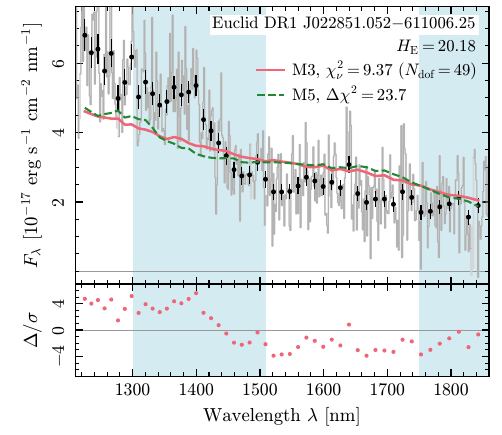}\hfill
  \includegraphics[width=0.33\textwidth]{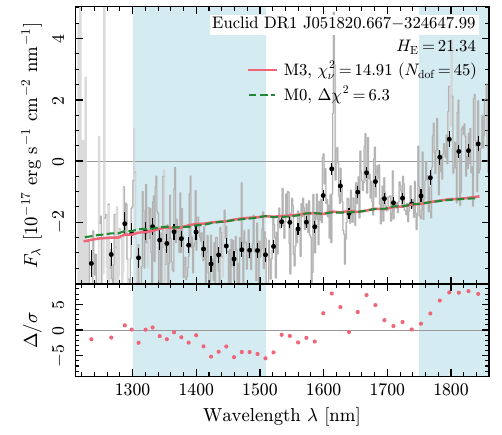}\\[2pt]
  \includegraphics[width=0.33\textwidth]{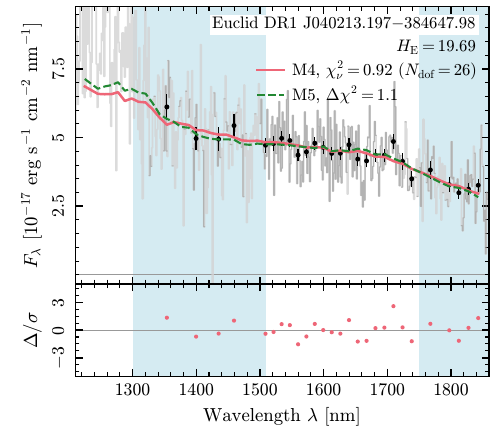}\hfill
  \includegraphics[width=0.33\textwidth]{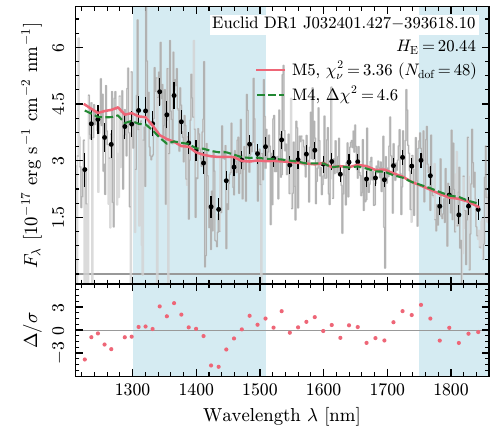}\hfill
  \includegraphics[width=0.33\textwidth]{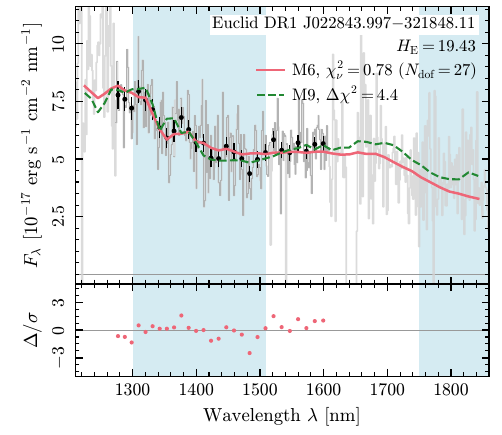}\\[2pt]
  \includegraphics[width=0.33\textwidth]{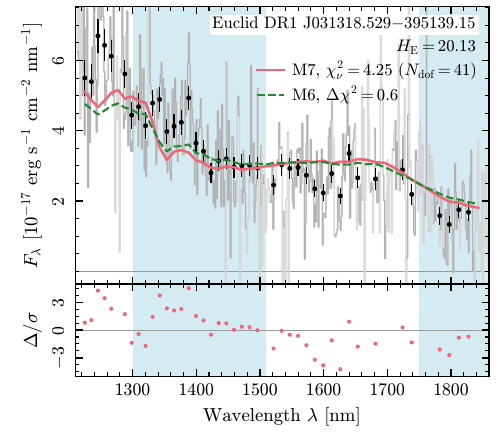}\hfill
  \includegraphics[width=0.33\textwidth]{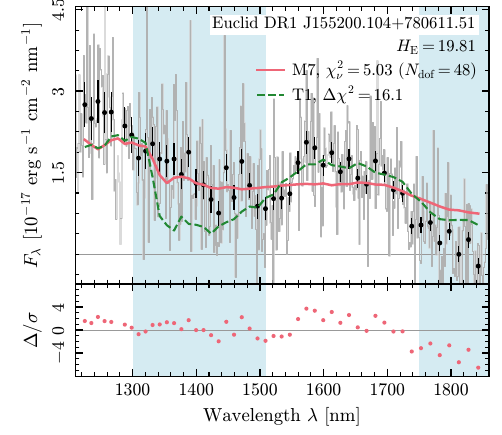}\hfill
  \includegraphics[width=0.33\textwidth]{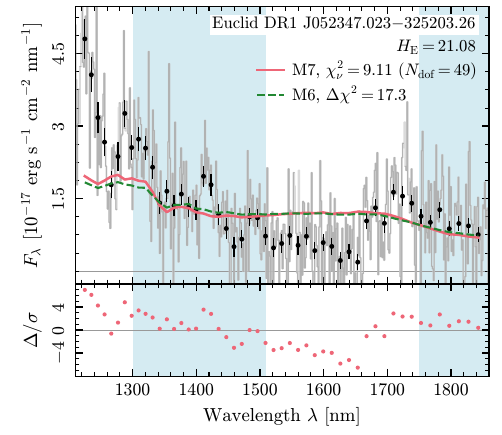}\\[2pt]
  \includegraphics[width=0.33\textwidth]{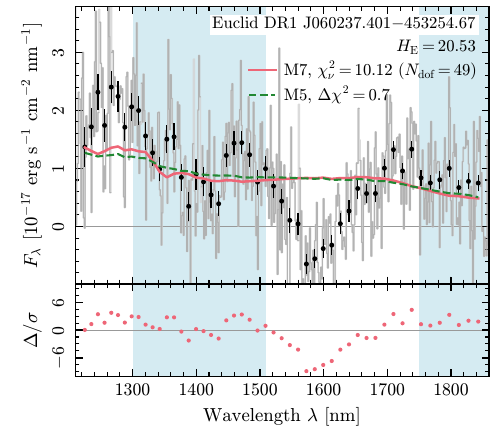}\hfill
  \includegraphics[width=0.33\textwidth]{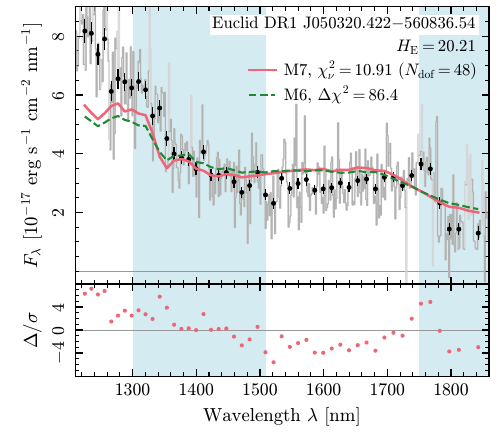}\hfill
  \includegraphics[width=0.33\textwidth]{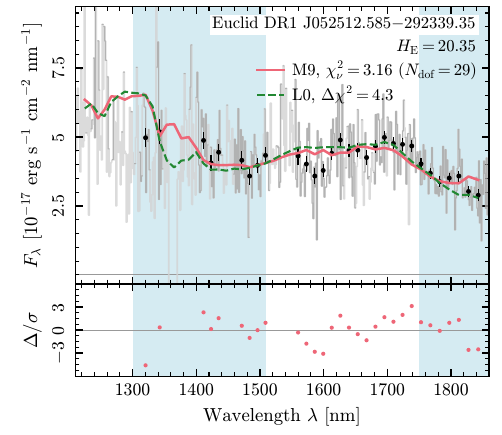}
  \caption{Continued.}
\end{figure*}
\begin{figure*}
  \addtocounter{figure}{-1}
  \renewcommand{\theHfigure}{\thefigure-cont}
  \centering
  \includegraphics[width=0.33\textwidth]{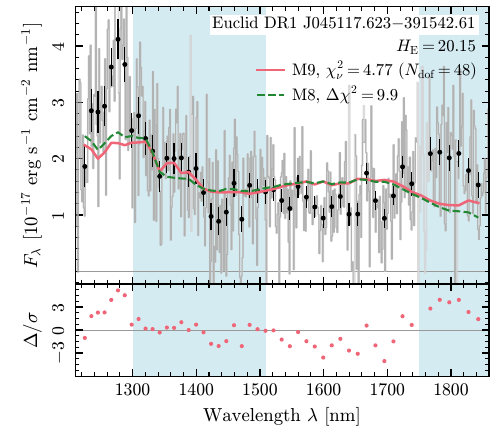}\hfill
  \includegraphics[width=0.33\textwidth]{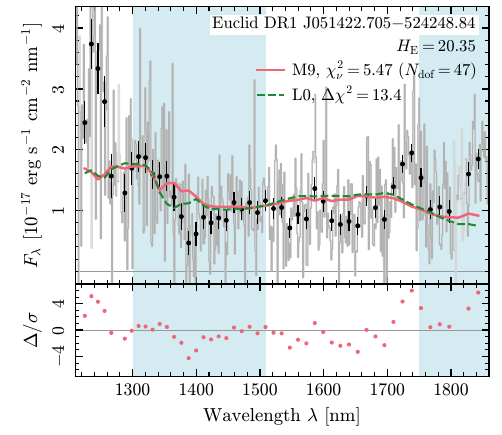}\hfill
  \includegraphics[width=0.33\textwidth]{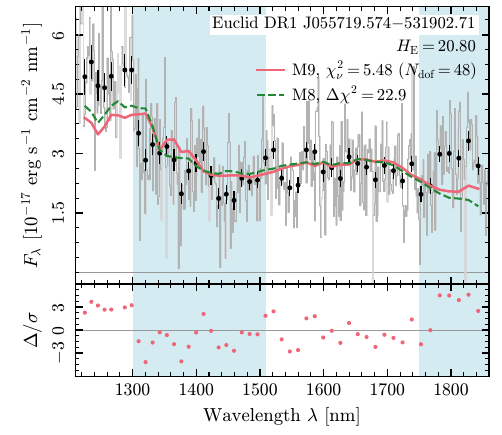}\\[2pt]
  \includegraphics[width=0.33\textwidth]{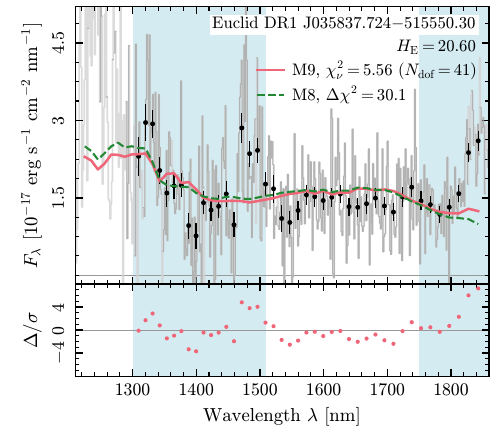}\hfill
  \includegraphics[width=0.33\textwidth]{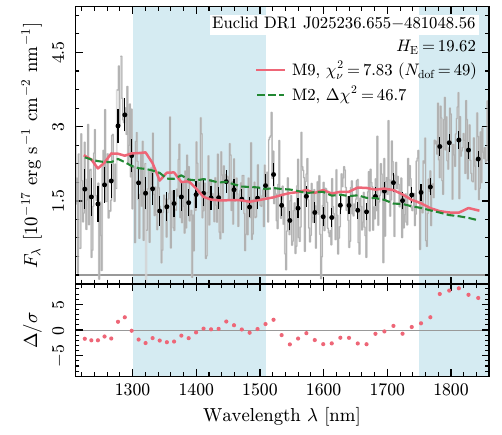}\hfill
  \includegraphics[width=0.33\textwidth]{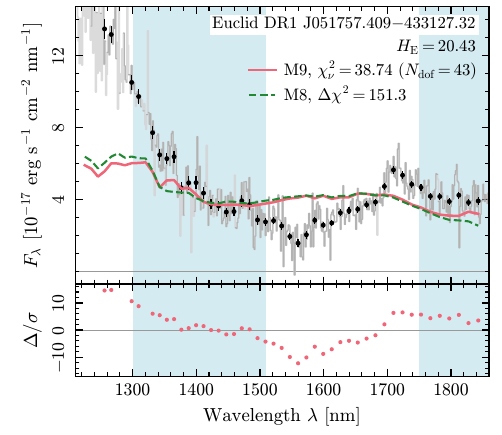}\\[2pt]
  \includegraphics[width=0.33\textwidth]{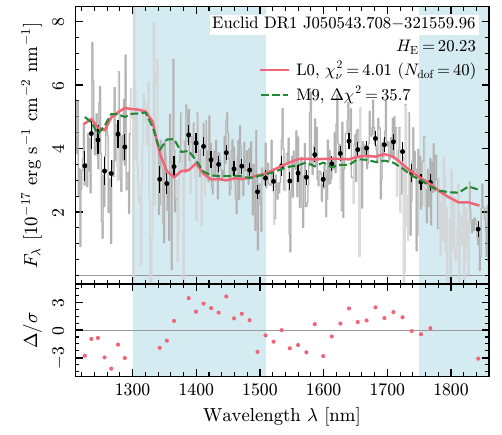}\hfill
  \includegraphics[width=0.33\textwidth]{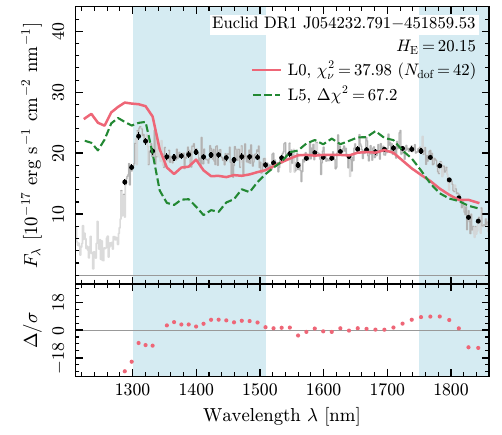}\hfill
  \includegraphics[width=0.33\textwidth]{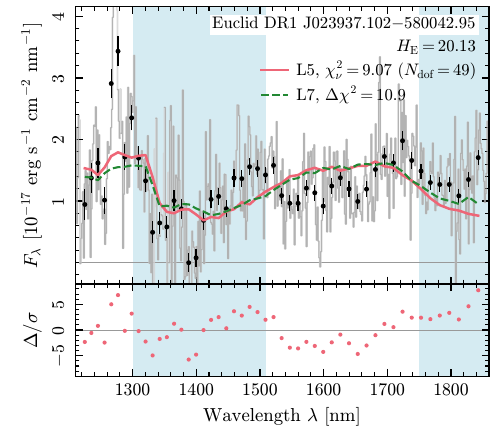}\\[2pt]
  \includegraphics[width=0.33\textwidth]{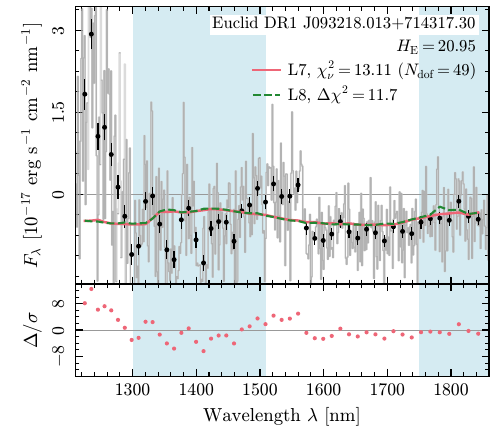}\hfill
  \includegraphics[width=0.33\textwidth]{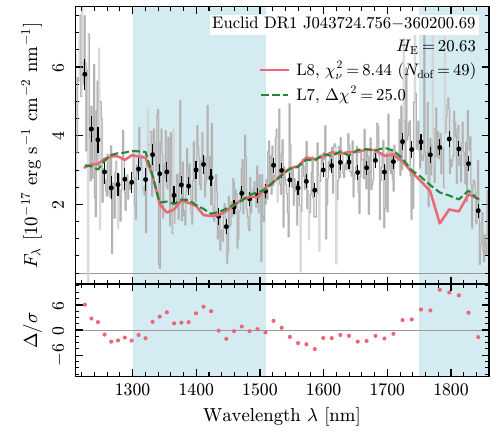}\hfill
  \includegraphics[width=0.33\textwidth]{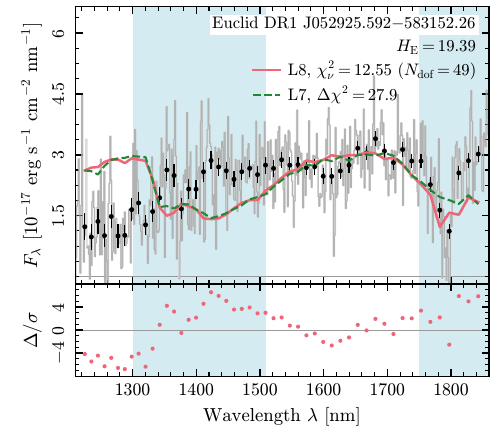}
  \caption{Continued.}
\end{figure*}
\begin{figure*}
  \addtocounter{figure}{-1}
  \renewcommand{\theHfigure}{\thefigure-cont}
  \centering
  \includegraphics[width=0.33\textwidth]{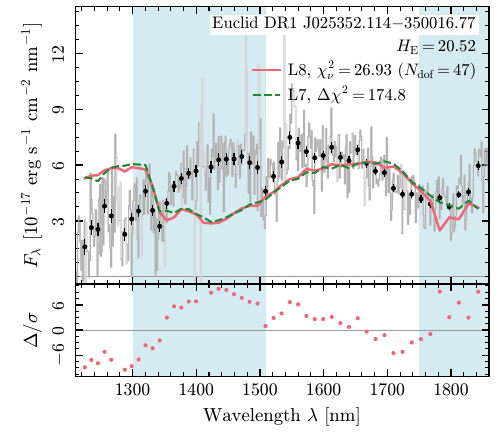}\hfill
  \includegraphics[width=0.33\textwidth]{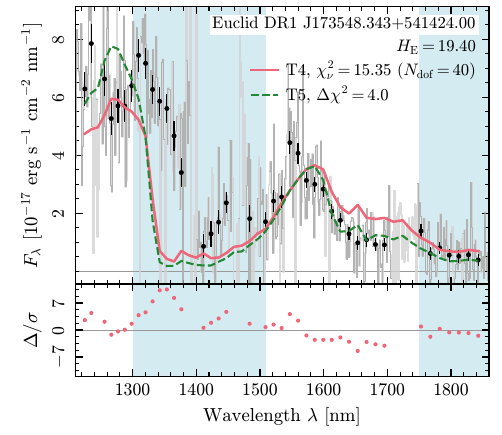}\hfill
  \includegraphics[width=0.33\textwidth]{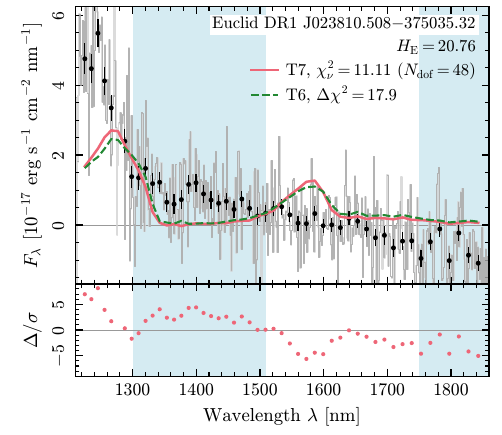}\\[2pt]
  \includegraphics[width=0.33\textwidth]{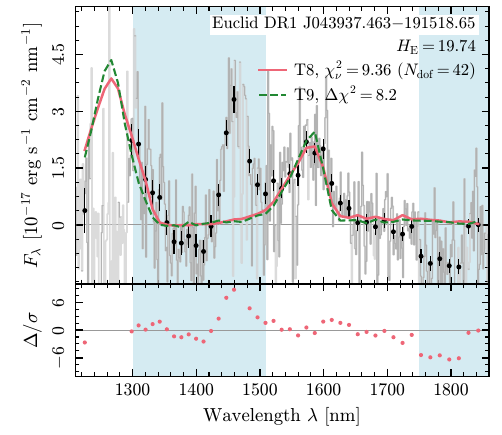}\hfill

  \caption{Continued.}
\end{figure*}

\end{appendix}

\label{LastPage}

\end{document}